\documentstyle[aps,epsf]{revtex}
\newcommand{\zh}{\mathbf}
\newcommand{\real}{\mathop{\rm Re }\nolimits}
\newcommand{\image}{\mathop{\rm Im }\nolimits}
\newcommand{\ordnung}{\mathop{  \rm O  }\nolimits}

\newcommand{\ground}{\mathop{ A }\nolimits}

\newcommand{\omegares}{\omega^{\text{res}}}

\newcommand{\oee}{\mathop{\varepsilon}\nolimits}
\newcommand{\mee}{\mathop{E}\nolimits}
\newcommand{\Eexch}{E_{\mbox{x}}}

\begin{document}
\draft
\bibliographystyle{revtex}
\title{Calculation of quasi-degenerate energy levels of
       two-electron ions}
\author{O.~Yu.~Andreev,${}^1$ L.~N.~Labzowsky,${}^{1,2}$
        G.~Plunien,${}^3$ and G.~Soff${}^3$}
\address{${}^1$ {V.~A.~Fock~Institute of Physics,
             St.~Petersburg State University, Ulyanovskaya 1,
             198504,
             Petrodvorets, St.~Petersburg, Russia}}
\address{${}^2$ {Petersburg Nuclear Physics Institute,
             188300,
             Gatchina, St.~Petersburg, Russia}}
\address{${}^3$ {Institut f\"ur Theoretische Physik,
             Technische Universit\"at Dresden,
             Mommsenstra{\ss}e 13, D-01062, Dresden, Germany}}
\date{\today}
\maketitle
\begin{abstract}
Accurate QED calculations of the interelectron interaction
corrections for the
$(1s2p)2\,{}^1\! P_1$, $(1s2p)2\,{}^3\! P_1$
two-electron configurations for ions with nuclear charge numbers
$10\le Z \le 92$
are performed within the line profile approach.
Total energies of these configurations are evaluated.
Employing the fully relativistic treatment based on the
\mbox{$j$--$j$} coupling scheme
these energy levels become quasi-degenerate in the region
$Z\le 40$.
To treat such states within the framework of QED
we utilize the line profile approach.
The calculations are performed within the Coulomb gauge.
\end{abstract}
\pacs{PACS number(s): 31.30.Jv, 31.10.+z}
\section{Introduction}
\label{introduction}
To provide
accurate quantum-electrodynamical (QED) evaluations of energy
levels for two- and three-electron configurations of
highly charged ions (HCI) become now an urgent problem in
atomic physics.
This can be explained by the growing number of experimental
data and the necessity to use the energy levels for the evaluation
of important
characteristics of HCI, such as e.g. transition probabilities
and recombination cross sections.

\par
In the past an approximate relativistic approach based on
variational non-relativistic wave functions
has been used for
evaluating energy levels
\cite{drake88}.
Numerous theoretical results for few-electron ions have been obtained
within the framework of fully relativistic many-body
perturbation theory (RMBPT) and relativistic all-order
many-body theory (AO)
\cite{plante94}.
However, rigorous QED results, which allow for a consequent
order-by-order improvement of the accuracy of the calculations
become more and more desirable.

\par
The approximation of non-interacting electrons is commonly
employed as a zeroth-order approximation in QED perturbation
theory for HCI in the absence of a quasi-degeneracy of levels.
Accordingly, within the zeroth-order the energy of
the few-electron configuration appears
as the sum of the Dirac eigenvalues for the occupied
one-electron states.
One-electron corrections (termed here as generalized Lamb shift)
include QED radiative corrections, such as
the electron self energy (SE) and vacuum polarization (VP)
as well as nuclear corrections, i.e.,
nuclear size (NS), nuclear
recoil (NR) and nuclear polarization (NP), respectively.
Few years ago a status report of one-electron energy corrections
has been presented in detail in
\cite{mohr98}.
Since then considerable progress
concerning
the evaluation of
higher-order self-energy corrections
has been made
\cite{yerokhin03}.

\par
The dominant two-electron contribution is due to
the interelectron interaction.
Ab initio QED results for the first-order interelectron
interaction in two-electron ions are known from
Ref.~\cite{klimchitskaya71}
(see also
\cite{labzowsky93b}).
The higher-order corrections are much more intricate.
Complete QED calculations of the second-order interelectron
interaction have been accomplished for the ground state and
for non-degenerate
low-lying excited states of He-like and Li-like ions
\cite{blundell93,lindgren95,yerokhin00%
,mohr00,andreev01,asen02,andreev03}.

\par
The other important two-electron corrections are
the screened self-energy and vacuum-polarization
corrections, which have been evaluated in
\cite{indelicato01,artemyev00,yerokhin99,artemyev99}
for non-degenerate
two- and three-electron configurations in HCI.

\par
Various general bound-state QED approaches have been employed
for the derivation of the energy corrections in HCI.
The one most commonly used is the adiabatic S-matrix approach,
developed by Gell-Mann and Low
\cite{gellmann51},
generalized by Sucher
\cite{sucher57}
and first applied to bound-state QED corrections in
Ref.~\cite{labzowsky70}
(see also
\cite{klimchitskaya71}).
The other one is the Green's function approach, first developed in
\cite{braun77}
and now applied frequently in a more sophisticated form of the two-time
Green's function method
\cite{shabaev90,shabaev93,shabaev02}.
Finally, the line profile approach (LPA) is utilized for the same
purpose
\cite{labzowsky93karasiev,labzowsky02annals}.
In our previous studies on the QED
theory of interelectron interaction in HCI
\cite{andreev01,andreev03}
this approach has been applied.

\par
In this paper we generalize the line profile approach to the case
of quasi-degenerate electronic states in two-electron HCI.
This problem arises, when a complete QED treatment including the
relativistic
\mbox{$j$--$j$}
coupling scheme is applied to the fine-structure multiplets
of systems
with intermediate nuclear charge numbers $Z$.
Some components of the
multiplet with equal relativistic quantum numbers turn out
to be close enough to each other
(the $(1s2p)2\,{}^1\! P_1$, $(1s2p)2\,{}^3\! P_1$
levels are the standard example).

\par
Up to now the QED theory of the quasi-degenerate levels was
considered only within the two-time Green's function method
for the self-energy screening corrections
(without any numerical applications)
\cite{lebigot01},
for vacuum-polarization screening corrections
\cite{artemyev00}
and within the covariant evolution-operator procedure
\cite{lindgren01}
for the second-order interelectron interaction.
Numerical calculations for two-electron ions with
$Z=10,18$
\cite{lindgren01}
are so far the only example where bound-state QED
has been applied
to the evaluation of the energy of quasi-degenerate
levels taking into account the interelectron interaction
up to second order.

\par
In this work we present an extension of the line profile approach,
which is suitable for the evaluation of energies of any
number of the non-degenerate or quasi-degenerate levels.
The interelectron interaction corrections up to first- and
second-order of QED perturbation theory are evaluated for the
$(1s2p) 2\,{}^1\! P_1$
and
$(1s2p) 2\,{}^3\! P_1$
levels in a wide range of $Z$ values.

\section{Line profile approach}
\label{lpa}
The problem of the natural line profile in atomic physics was
considered first in terms of quantum mechanics by Weisskopf and
Wigner
\cite{weisskopf30}.
In terms of modern QED it was first formulated for one-electron
atoms by Low
\cite{low52}.
In
\cite{low52}
the appearance of the Lorentz profile
in the resonance approximation
within the framework of QED was described and nonresonant
corrections were estimated.
Later the line profile QED theory was modified also for
two-electron atoms
\cite{labzowsky83}
(see also
\cite{labzowsky93b,labzowsky93only})
and applied to the theory of overlapping resonances in
two-electron HCI
\cite{gorshkov89,karasiev92}.
Another application was devoted to the theory of nonresonant
corrections
\cite{labzowsky94,labzowsky97goidenko}.

\par
It was found in
\cite{labzowsky93karasiev},
that the LPA provides a convenient tool for
calculating energy corrections.
Moreover,
it clearly determines the limit up to which the concept of the
energy of the excited states has a physical meaning -- that is the
resonance approximation.
The exact theoretical values for the energy of the excited states
defined, by the poles in the Green's function,
can be directly compared with
measurable quantities only within the resonance approximation, where
the line profile is described by the two parameters:
energy $E$ and width $\Gamma$.
Beyond this approximation the evaluation of
$E$ and $\Gamma$
should be replaced by the evaluation of the line profile for the
particular process.
Moreover, in the case of two-electron atoms
the line profile approach was found to be very efficient
for the evaluation of the reference state correction
(reducible part of Feynman graphs)
for two-electron atoms
\cite{labzowsky98}.

\subsection{Line profile approach for one-electron ions}
\label{lpa1}
\begin{figure}
\centerline{ \mbox{ \epsfxsize=0.20\textwidth \epsffile{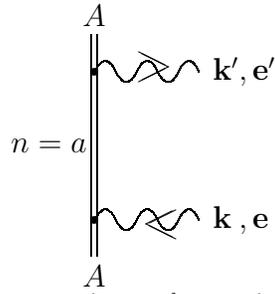} }}
\caption{
The lowest-order amplitude of the photon scattering on the atomic
electron within the resonance approximation.
The double solid line corresponds to bound electrons in the field
of the nucleus.
The wavy lines with arrows denote the absorption or the emission
of a photon with momentum
${\zh{k}}$
and polarization
$\zh{e}$.
}
\label{figure06}
\end{figure}

Consider the simplest process of photon scattering on
a one-electron ion which is assumed to be in
its ground state
$\ground$
(Fig. \ref{figure06}).
Using the standard Feynman rules
for  bound-electron QED
\cite{labzowsky93b}
yields
the expression for the S-matrix element
\begin{eqnarray}
S_{\ground}^{(2)}
\label{rsco2x031212x3}
\label{lpaEq1}
&=&
(-ie)^2 \int d^4x_{u} \, d^4x_{d}\,
\bar{\psi}_{\ground}(x_u) \gamma^{\mu_u}
S(x_u, x_d)
\gamma^{\mu_d}\psi_{\ground}(x_d)
\nonumber\\
&&
\times
A^{\ast(k',\lambda')}_{\mu_u}(x_u)
A^{(k,\lambda)}_{\mu_d}(x_d)
\,,
\end{eqnarray}
where
$\psi_{\ground}(x)=
\psi_{\ground}({\zh r})e^{-i\oee_{\ground}t}$
is the wave function of the electron in the ground state,
$\gamma^{\mu}$ is the Dirac matrix
together with the electron propagator
\begin{eqnarray}
S(x_1,x_2)
\label{proppsi}
\label{propagatorelectron}
&=&
\frac{i}{2\pi}
\int\nolimits_{-\infty}^{\infty}d\omega\,e^{-i\omega(t_1-t_2)}
\sum_n
\frac{\psi_n({{\zh r}_1})\bar{\psi}_n({{\zh r}_2})}
     {\omega-\oee_n(1-i0)}
\,.
\end{eqnarray}
$A^{(k,\lambda)}_{\mu}(x)$
denotes the vector potential of the electromagnetic
field (photon wave function).
The notations
$x_u=(t_u,{\zh r}_u)$ and
$x_d=(t_d,{\zh r}_d)$ indicate ``up'' and ``down''
vertex coordinates, respectively.

\par
Insertion of the expressions for the electron propagator and
the photon wave function yields
\begin{eqnarray}
S_{\ground}^{(2)}
\label{lpaEq2}
&=&
(-ie)^2 \int dt_u \, d^{3}{\zh r}_u \, dt_d \,
d^{3}{\zh r}_d \, d\omega_{n} \,\,
[\bar{\psi}_{\ground}({\zh r}_u)\gamma^{\mu_u}
A^{\ast(k',\lambda')}_{\mu_u}({\zh r}_u)]
\nonumber\\
&&
\times
e^{it_u(\oee_{\ground}+\omega')}
e^{-i\omega_{n}(t_u-t_d)}
\frac{i}{2\pi}
\sum\limits_{n}
\frac{\psi_{n}({\zh r}_u)\bar{\psi}_{n}({\zh r}_d)}
{\omega_{n}-\oee_{n}(1-i0)}
e^{-it_d(\oee_{\ground}+\omega)}
\nonumber\\
&&
\times
[\gamma^{\mu_d}A^{(k,\lambda)}_{\mu_d}({\zh r}_d)
\psi_{\ground}({\zh r}_d)]
\,.
\end{eqnarray}
Here
$\omega=|{\zh k}|$
and
$\omega'=|{\zh k}'|$
are frequencies of the absorbed and emitted photons,
respectively,
${\zh k},{\zh k}'$ are the photon momenta and
$\lambda,\lambda'$ denote the photon polarizations.
The summation over $n$ is extended over the entire Dirac spectrum
of electrons in the nuclear Coulomb field,
$\oee_{n}$
are the Dirac energy eigenvalues.
Integrating over time variables
($t_u, t_d$)
and abbreviating the expressions in the square brackets by
$\bar{\Phi}_{\ground}({\zh r}_u)$ and
$\Phi_{\ground}({\zh r}_d)$,
respectively,
we can write
\begin{eqnarray}
S_{\ground}^{(2)}
\label{lpaEq3}
&=&
(-ie)^2 (2\pi)^2 \int  d^{3}{\zh r}_u  \, d^{3}{\zh r}_d \,
d\omega_{n} \,\,
\bar{\Phi}_{\ground}({\zh r}_u)
\nonumber\\
&&
\times
\delta(\omega_{n}-\oee_{\ground}-\omega')
\frac{i}{2\pi}
\sum\limits_{n}
\frac{\psi_{n}({\zh r}_u)\bar{\psi}_{n}({\zh r}_d)}
{\omega_{n}-\oee_{n}(1-i0)}
\delta(\oee_{\ground}+\omega-\omega_{n})
\nonumber\\
&&
\times
\Phi_{\ground}({\zh r}_d)
\,.
\end{eqnarray}
The function $\Phi_{\ground}({\zh r})$
can be considered as a vertex function,
which describes the absorption of a photon by an
electron in its ground state.
Below, we will formulate the resonance approximation, where
we can define the energy and width of the level which
have the general meaning independent of
the features of the considered scattering process.
Hence, the energy and the width will not depend on
the function
$\Phi_{\ground}({\zh r})$
and thus
we may consider the function
$\Phi_{\ground}({\zh r})$
as arbitrary.
In particular,
it can account for the interaction with
the free electromagnetic field (radiative corrections).

\par
Let us introduce in
Eq.~(\ref{lpaEq3})
the matrix
\begin{eqnarray}
T_{n \ground}
&=&
(-e) \int   d^{3}{\zh r}  \,\,
\bar{\psi}_{n}({\zh r})
\Phi_{\ground}({\zh r})
\,
\end{eqnarray}
and integrate over the frequency
$\omega_{n}$.
Employing the definition of the transition amplitude $U$ via
\begin{eqnarray}
S
\label{defU}
&=&
-2\pi i
\,
\delta(\omega-\omega')
\,
U
\,
\end{eqnarray}
we derive the expression for the amplitude
\begin{eqnarray}
U_{\ground}^{(2)}
\label{rsco2}
&=&
\sum_{n}
\frac{
  T^{\ast}_{\ground n}
  T_{n \ground}
  }{\omega-\oee_n + \oee_{\ground}}
\,.
\end{eqnarray}
We will consider the resonance case,
when the frequency
$\omega$
is close to the value
$\omegares=\oee_{a}-\oee_{\ground}+\ordnung(\alpha)$,
where
$a$
labels one of the exited states of an ion.
In the resonance approximation we have to retain in
Eq.~(\ref{rsco2})
only the dominant term
with $n=a$
in the sum over
$n$, i.e.,
\begin{eqnarray}
U_{\ground a}^{(2)}
\label{plaresonance}
&=&
\frac{
  T^{\ast}_{\ground a}
  T_{a \ground}
  }{\omega-\oee_a + \oee_{\ground}}
=
T^{\ast}D^{-1}T
\,.
\end{eqnarray}
In order to simplify the application of the line profile approach
to the many-electron ions
we introduce the abbreviated notations
\begin{eqnarray}
\label{y1x031212x4}
T
&=&
T_{a \ground}
\,,
\\
D
&=&
{\omega - V^{(0)}+\oee_{\ground}}
\,,
\\
V^{(0)}
&=&
\oee_{a}
\,.
\end{eqnarray}
Notice, that the function $T$
describes the process of scattering.

\par
\begin{figure}
\centerline{ \mbox{ \epsfxsize=0.19\textwidth \epsffile{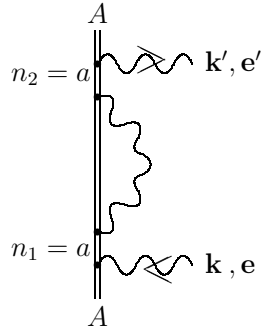} }}
\caption{
First-order self-energy insertion into the photon-scattering graph
within the resonance approximation.
}
\label{figure07}
\end{figure}

To obtain the Lorentz contour one has to insert the electron
self-energy part in the internal electron line in
Fig.~\ref{figure06}.
For simplicity we neglect the vacuum-polarization part.
To the lowest order this leads to the graph depicted in
Fig.~\ref{figure07}
and the corresponding expression for the scattering amplitude
evaluated within the resonance approximation takes the form
\begin{eqnarray}
\label{y1x031212x5}
U_{\ground a}^{(4)}
&=&
U_{\ground a}^{(2)}\quad
\frac{V^{(1)}(\omega)}{\omega-\oee_{a}+\oee_{\ground}}
=
T^{\ast}D^{-1}\left[V^{(1)}(\omega)D^{-1}\right]T
\end{eqnarray}
with
\begin{eqnarray}
V^{(1)}(\omega)
\label{v1se}
&=&
e^2
\,
\left(\hat{\Sigma}_{\rm R}(\omega+\oee_{\ground})\right)_{aa}
\,.
\end{eqnarray}
Here
$\hat{\Sigma}_{\rm R}(\omega)$
is the renormalized electron self-energy operator.
The upper index at the function $V$
indicates the order of perturbation theory with respect to
powers of the fine structure constant $\alpha$
for the graphs contributing to this function.
Repeating these insertions in
higher orders we can compose
a geometric progression with the $l$-th term
\begin{eqnarray}
Q_{l}
&=&
U^{(2)}_{\ground a}
\left[\frac{V^{(1)}(\omega)}{\omega-\oee_{a}+
\oee_{\ground}}\right]^{l}
=
T^{\ast}D^{-1}\left[V^{(1)}(\omega)D^{-1}\right]^l T
\,.
\end{eqnarray}
The resulting geometric progression is convergent for any
$\omega$
except for values within the interval
{$\omega \in [\oee_{a}-\oee_{\ground}-|V^{(1)}|,
\oee_{a}-\oee_{\ground}+|V^{(1)}|]$}
close to the position of the resonance.
Applying the formula for a convergent geometric
progression one derives
\begin{eqnarray}
U_{\ground a}
\label{rscse}
\label{converGeometrProgr1}
\label{converGeometrProgr2}
&=&
\sum\limits^{\infty}_{l=0}
T^{\ast}D^{-1}\left[V^{(1)}(\omega)D^{-1}\right]^l T
\nonumber
\\
\label{converGeometrProgr3}
&=&
\frac{T^{\ast}T}{D-V^{(1)}(\omega)}
=
\frac{T^{\ast}T}{\omega+\oee_{\ground}-V^{(0)}-V^{(1)}(\omega)}
\,.
\end{eqnarray}
Hence, the resonance is shifted into the complex plane and
Eq.~(\ref{converGeometrProgr3})
is defined for all $\omega$ values on the real axis.
Eq.~(\ref{converGeometrProgr3})
presents the analytic continuation of the expansion
$\sum^{\infty}_{l=0} Q_l$
to the entire complex plane.

\par
Taking the square modulus of the amplitude
(\ref{rscse}),
integrating over the directions of absorbed and emitted
photons
and summing over the polarizations we obtain
the Lorentz profile
for the absorption probability
\begin{eqnarray}
d W(\omega)
\label{rscw}
&=&
\frac{1}{2\pi}
\frac{\Gamma_{a \ground}}
{(\omega+\oee_{\ground}
-V^{(0)}-{\real}\left\{V^{(1)}(\omega)\right\})^2+
({\image}\left\{V^{(1)}(\omega)\right\})^{2}
}
\,
d\omega
\,.
\end{eqnarray}
Here
$d W(\omega)$
is the probability
for the absorption of a photon
within the frequency interval
$\omega$,
$\omega+d\omega$
and
$\Gamma_{a \ground}$
is the partial width of the level
$a$,
associated with the transition
$a\to \ground$.

\par
Taking into account the correction depicted in
Fig.~\ref{figure07}
we improve the position of the resonance
\begin{eqnarray}
\omegares
\label{rscse2lim651238}
&=&
-\oee_{\ground}+V^{(0)}
+
{\real}\left\{V^{(1)}(\oee_{a}-\oee_{\ground})\right\}
+
\ordnung(\alpha^2)
\,.
\end{eqnarray}
Formula
(\ref{rscw})
defines the line profile of the process of scattering.
Within the resonance approximation the line profile
can be described by a Lorentz contour
which is characterized by two parameters:
the position of the resonance and the width.
We define the energy shift
for the state $a$
as the shift of the resonance.
The energy of the state $a$ is
\begin{eqnarray}
E
\label{lpaDefE}
&=&
\omegares+\oee_{\ground}
=
V^{(0)}
+
{\real}\left\{V^{(1)}(\oee_{a}-\oee_{\ground})\right\}
+
\ordnung(\alpha^2)
\,
\end{eqnarray}
and the width of the level as
the width of the corresponding Lorentz contour at the position
of the resonance
\begin{eqnarray}
\Gamma
\label{lpaDefGamma}
&=&
-2\,\,
{\image}\left\{V(\omegares)\right\}
=
-2\,\,
{\image}\left\{V^{(1)}(\oee_{a}-\oee_{\ground})\right\}
+
\ordnung(\alpha^2)
\,.
\end{eqnarray}
We note that the energy $E$ of the level and
its width $\Gamma$ defined
in the framework of the resonance approximation do
not depend upon the function $T$ (or $\Phi_{\ground}$)
and, therefore, they do not depend upon the type of the
scattering process.
For example, we are free to consider not only the scattering of
a photon but of some other particle as well,
which couples to electrons.
Going beyond the resonance approximation
the line profile can no longer be described by a Lorentz contour
and, consequently, the energy level can not be characterized
only by two parameters
$E$ and $\Gamma$.
In this case
the evaluation of the energy levels should be replaced by
the evaluation of the particular line profiles which, in general,
depend upon the features of scattering process under consideration.

\par
The real part of the matrix element
$\left({\hat\Sigma}_{\rm R}(\oee_{a})\right)_{aa}$
describes the lowest-order contribution to the Lamb shift and the
imaginary part, which is finite and not subject to renormalization,
represents the total radiative (single-quantum) width of the level
$a$:
\begin{eqnarray}
\Delta E^{\rm SE}_{a}
\label{xxxx2}
&=&
\left(\hat{\Sigma}_{\rm R}(\oee_{a})\right)_{aa}
=
L^{\rm SE}_{a}
-\frac{i}{2}
\,
\Gamma_{a}
\,.
\end{eqnarray}
The other contribution to the lowest-order Lamb shift
$L_{a}^{\rm VP}$
originates from the vacuum polarization.
This correction does not contribute to the width
$\Gamma_{a}$
\cite{labzowsky93b}.

\par
\begin{figure}
\centerline{ \mbox{ \epsfxsize=0.06\textwidth \epsffile{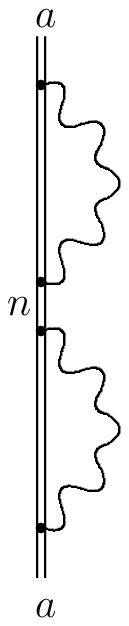} }}
\caption{
The second-order electron self-energy correction
(so called SESE loop-after-loop)
that gives rise to the correction in
Eq. (\ref{seseo6nr}).
}
\label{figure08}
\end{figure}
\begin{figure}
\centerline{ \mbox{ \epsfxsize=0.15\textwidth \epsffile{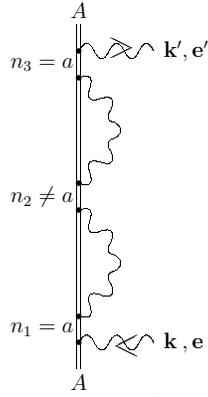} }}
\caption{
Feynman graph representing the higher order electron self-energy
correction within the line profile approach
(SESE, loop-after-loop, irreducible).
}
\label{figure09}
\end{figure}

Studying the higher-order Lamb shift in one-electron
atoms within the line profile approach,
we have to account for the Feynman graph depicted in
Fig.~\ref{figure08}.
For reason of simplicity, we will not consider
the other second-order graphs.
In the case $n_1=n_3=a$ and $n_2\ne a$ the graph in
Fig.~\ref{figure09}
can be viewed as a second-order self-energy insertion
(loop-after-loop, irreducible part)
in the graph
Fig.~\ref{figure06}
within the resonance approximation.
We derive the following expression for the scattering amplitude
\begin{eqnarray}
\label{y2x021212x1}
U_{\ground a}^{(6)}
&=&
U_{\ground a}^{(2)}\quad
\frac{V^{(2)}(\omega)}{\omega-\oee_{a}+\oee_{\ground}}
=
T^{\ast}D^{-1}\left[V^{(2)}(\omega)D^{-1}\right]T
\,,
\end{eqnarray}
where
\begin{eqnarray}
V^{(2)}(\omega)
\label{v2e1ph2}
&=&
e^4
\sum_{n\ne a}
\frac{
  \left(
  \hat{\Sigma}_{\rm R}(\omega+\oee_{\ground})
  \right)_{an}
  \left(
  \hat{\Sigma}_{\rm R}(\omega+\oee_{\ground})
  \right)_{n a}
}
{\oee_{\ground} - \oee_{n}+\omega}
\,.
\end{eqnarray}
Note, that the singular term
$n=a$
is not included here by definition.
This term was taken into account in the geometric
progression described above and represents
exactly the second term of this progression.
Repeating the evaluations leading to
Eq. (\ref{rscw})
with
\begin{eqnarray}
Q_{l}
&=&
U^{(2)}_{\ground a}
\left(\frac{V^{(1)}(\omega)+V^{(2)}(\omega)}
           {\omega-\oee_{a}+\oee_{\ground}}\right)^{l}
\nonumber
\\
&=&
T^{\ast}D^{-1}\left[(V^{(1)}(\omega)
+V^{(2)}(\omega))D^{-1}\right]^l T
\,,
\end{eqnarray}
we obtain the improved resonance condition
\begin{eqnarray}
\label{o2resequ}
V^{(0)}
+
{\real}\{\,
V^{(1)}(\omegares)
+
V^{(2)}(\omegares)
\,\}
+
\ordnung(\alpha^3)
-\oee_{\ground}
-\omegares
&=&
0
\,.
\end{eqnarray}
Solving this equation for $\omegares$
up to terms $\ordnung(\alpha^3)$ yields
\begin{eqnarray}
\label{seseo6nr}
\omegares
&=&
-\oee_{\ground}
+
V^{(0)}
+
{\real}\{\, V^{(1)}(\oee_{a}-\oee_{\ground})
+
V^{(2)}(\oee_{a}-\oee_{\ground})
\nonumber\\
&&
+
V^{(1)}(\oee_{a}-\oee_{\ground})
\left[\frac{\partial V^{(1)}(\omega)}{\partial\omega}
\right]{\vphantom{\frac{8}{8}}}_{
\textstyle \omega=\oee_{a}-\oee_{\ground}}
\,\}
+
\ordnung(\alpha^3)
\,.
\end{eqnarray}
The term
$V^{(2)}(\oee_{a}-\oee_{\ground})$
is the contribution of the irreducible part of the graph in
Fig. \ref{figure08}.
The derivative term corresponds to the
reference state ($n=a$) correction.
In particular, it coincides with the reference state
correction that arises from the Feynman graph in
Fig. \ref{figure08}
after application of the adiabatic S-matrix method
\cite{labzowsky96}.
The other second-order electron self-energy (SESE)
corrections are irreducible
\cite{labzowsky96}.

\subsection{Line profile approach for many-electron ions
            (non-degenerate energy level)}
\label{lpa2}
As in the one-electron case we will consider the process of
photon scattering on an ion which is assumed to be in
its lowest (ground) state.
Investigating a non-degenerate energy level associated
with a configuration
containing at least one $1s$-electron
(such as
$(1s2s)2 {}^1\!S_{0}$,
$(1s2p_{1/2})2 {}^3\!P_{0}$,
$(1s2s)2 {}^3\!S_{1}$,
$(1s)^22s_{1/2}$,
$(1s)^22p_{1/2}$,
etc)
we can represent the wave function of the ground state
via a proper combination of one-electron Dirac wave functions.
A procedure based on this approach has been accomplished in
\cite{andreev01,andreev03}.
However, for the investigation of quasi-degenerate levels
or a doubly excited level
the interelectron interaction corrections 
have to be taken into account
in the wave function of the ground state.
Here we restrict ourselves to two-electron ions.
The generalization to N-electron ions will be presented at
the end of this section.

\par
\begin{figure}
\centerline{ \mbox{ \epsfxsize=0.18\textwidth \epsffile{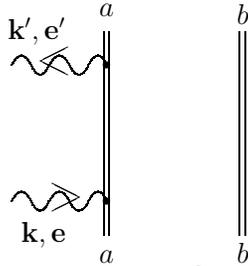} }}
\caption{
Lowest-order amplitudes for photon scattering on
a two-electron ion in its ground state
$\ground$
within the resonance approximation.
The ground state $\ground$ is represented in terms
of non-interacting Dirac electrons.
}
\label{figure27}
\end{figure}

In the one-electron case we introduced the function
$\Phi_{\ground}$
describing the process of scattering under consideration.
To introduce such a function for the two-electron system
we may consider first the simplest process of photon scattering
on a two-electron ion
disregarding the interelectron interaction corrections to
the initial (ground) state.
This process is depicted in
Fig.~\ref{figure27},
where the ground state is represented by two non-interacting
electrons one of which
absorbs (or emits) the photon.
Accordingly, the ground-state wave function is given by
\begin{eqnarray}
\Psi_{\ground}(x_1,x_2)
\label{detx1}
&=&
\frac{1}{2}\det\{\psi_{1s}(x_1)\psi_{1s}(x_2)\}
\,,
\\
\bar{\Psi}_{\ground}(x_1,x_2)
\label{detx2}
&=&
\frac{1}{2}\det\{\bar{\psi}_{1s}(x_1)\bar{\psi}_{1s}(x_2)\}
\,,
\end{eqnarray}
where
$\psi_{1s}(x_1)=\psi_{1s}({\zh r}_1)e^{-i\oee_{1s}t_1}$,
$\psi_{1s}(x_2)$
are the Dirac one-electron functions
with different projections of the total angular momentum.
The bar over the one-electron functions indicates
the Dirac conjugation.

\par
The S-matrix element corresponding to the graph in
Fig.~\ref{figure27}
can be written as
\begin{eqnarray}
S_{\ground}^{(2)}
\label{lpaEq3x1old2}
&=&
(-ie)^2
\frac{1}{2}
\nonumber\\
&&
\times
\left[
\int d^{4}{x_{u_1}} \, d^{4}{x_{u_2}}
\, d^{4}{x_{d_1}} \, d^{4}{x_{d_2}}
\delta^{3}({\zh r}_{u_2}-{\zh r}_{d_2})
\delta(t_{u_2})\delta(t_{d_2})
\, d{\omega}_{n_1}\,\,
\bar{\Psi}_{\ground}({x_{u_1}},{x_{u_2}})
\vphantom{
\sum\limits_{n_2}
\frac{\psi_{n_2}({{\zh r}_{u_2}})\bar{\psi}_{n_2}({{\zh r}_{d_2}})}
{\omega_{n_2}-\oee_{n_2}(1-i0)}
}
\gamma^{\mu_{u_1}}A^{\ast(k',\lambda')}_{\mu_{u_1}}(x_{u_1})
\right.
\nonumber\\
&&
\times
\left.
e^{-i\omega_{n_1}(t_{u_1}-t_{d_1})}
\frac{i}{2\pi}
\sum\limits_{n_1}
\frac{\psi_{n_1}({{\zh r}_{u_1}})\bar{\psi}_{n_1}({{\zh r}_{d_1}})}
{\omega_{n_1}-\oee_{n_1}(1-i0)}
\gamma^{\mu_{d_1}}A^{(k,\lambda)}_{\mu_{d_1}}(x_{d_1})
\Psi_{\ground}({x_{d_1}},{x_{d_2}})
\right.
\nonumber\\
&&
+
\left.
\int d^{4}{x_{u_1}} \, d^{4}{x_{u_2}}
\,d^{4}{x_{d_1}}\, d^{4}{x_{d_2}}
\delta^{3}({\zh r}_{u_1}-{\zh r}_{d_1})
\delta(t_{u_1})\delta(t_{d_1})
\, d{\omega}_{n_2}\,\,
\bar{\Psi}_{\ground}({x_{u_1}},{x_{u_2}})
\gamma^{\mu_{u_2}}A^{\ast(k',\lambda')}_{\mu_{u_2}}(x_{u_2})
\right.
\nonumber\\
&&
\times
\left.
e^{-i\omega_{n_2}(t_{u_2}-t_{d_2})}
\frac{i}{2\pi}
\sum\limits_{n_2}
\frac{\psi_{n_2}({{\zh r}_{u_2}})\bar{\psi}_{n_2}({{\zh r}_{d_2}})}
{\omega_{n_2}-\oee_{n_2}(1-i0)}
\gamma^{\mu_{d_2}}A^{(k,\lambda)}_{\mu_{d_2}}(x_{d_2})
\Psi_{\ground}({x_{d_1}},{x_{d_2}})
\right]
\,.
\end{eqnarray}
In order to employ the functions
${\Psi}_{\ground}$,
$\bar{\Psi}_{\ground}$
we introduced additional integrations
$d^{4}x_{u_{1,2}} \, d^{4}x_{d_{1,2}}
\,\delta^{3}({\zh r}_{u_{1,2}}-{\zh r}_{d_{1,2}})
\,\delta(t_{u_{1,2}})\,\delta(t_{d_{1,2}})$.
The first and the second term in the square brackets
represents graphs,
where the photon is absorbed (emitted)
by the first or by the second electron, respectively.
Since the functions
$\Psi_{\ground}(x_1,x_2)$,
${\bar\Psi}_{\ground}(x_1,x_2)$
are antisymmetric it would be sufficient
to consider one of these terms only.

\par
As in the one-electron case
we will look for the position of the resonance
and employ the resonance approximation.
It implies the neglect of
the non-singular terms (evaluated at the resonance) in comparison
with singular ones.
The terms in the sum over $n_1$, $n_2$ in
Eq.~(\ref{lpaEq3x1old2})
may contain a singularity at the position of
the resonance only if
they correspond to the positive-energy part of the Dirac spectrum.
Accordingly, in
Eq.~(\ref{lpaEq3x1old2})
we can restrict to the terms with
$\oee_{n_{1}}>0$, $\oee_{n_{2}}>0$.

\par
Introducing the function
$\Phi_{\ground}(x_1,x_2)$
as
\begin{eqnarray}
\Phi_{\ground}(x_1,x_2)
&=&
\gamma^{\mu_1}A^{(k,\lambda)}_{\mu_1}(x_1)
\Psi_{\ground}(x_1,x_2)
\delta(t_1-t_2)
\,,
\\
\bar{\Phi}_{\ground}(x_1,x_2)
&=&
\bar{\Psi}_{\ground}(x_1,x_2)
\gamma^{\mu_1}A^{\ast(k',\lambda')}_{\mu_1}(x_1)
\delta(t_1-t_2)
\,,
\end{eqnarray}
we can write
\begin{eqnarray}
S_{\ground}^{(2)}
\label{lpaEq3x1}
&=&
(-ie)^2
\int  d^{4}{x_{u_1}}d^{4}{x_{u_2}}
 \, d^{4}{x_{d_1}}\, d^{4}{x_{d_2}}
\, d{\omega}_{n_1}\, d{\omega}_{n_2} \,\,
\bar{\Phi}_{\ground}({x_{u_1}},{x_{u_2}})
\nonumber\\
&&
\times
e^{-i\omega_{n_1}(t_{u_1}-t_{d_1})}
e^{-i\omega_{n_2}(t_{u_2}-t_{d_2})}
\nonumber\\
&&
\times
\frac{i}{2\pi}
\sum\limits_{n_1}
\frac{\psi_{n_1}({{\zh r}_{u_1}})\bar{\psi}_{n_1}({{\zh r}_{d_1}})}
{\omega_{n_1}-\oee_{n_1}(1-i0)}
\frac{i}{2\pi}
\sum\limits_{n_2}
\frac{\psi_{n_2}({{\zh r}_{u_2}})\bar{\psi}_{n_2}({{\zh r}_{d_2}})}
{\omega_{n_2}-\oee_{n_2}(1-i0)}
\nonumber\\
&&
\times
\Phi_{\ground}({x_{d_1}},{x_{d_2}})
\,.
\end{eqnarray}
Here we can employ the identity
\begin{eqnarray}
\frac{1}
{(\omega_{n_1}-\oee_{n_1}(1-i0))}
\frac{1}
{(\omega_{n_2}-\oee_{n_2}(1-i0))}
\label{polerevers}
&=&
\frac{2\pi}{i}
\frac{\delta(\omega_{n_1}-\oee_{n_1})}
{(\omega_{n_2}-\oee_{n_2}(1-i0))}
\nonumber\\
&&
+
\frac{-1}{(-\omega_{n_1}+\oee_{n_1}+i0\oee_{n_1})
(\omega_{n_2}-\oee_{n_2}(1-i0))}
\end{eqnarray}
which follows from the Sokhotsky formulas
\begin{eqnarray}
\frac{1}
{x+i0}
&=&
-i\pi\delta(x)
+
{\mathcal{P}}\frac{1}{x}
\,,\,\,\,\,\,\,\,\,\,
\frac{1}
{x-i0}
=
i\pi\delta(x)
+{\mathcal{P}}\frac{1}{x}
\,,\,\,\,\,\,\,\,\,\,
\frac{1}
{x+i0}
+
\frac{1}
{-x+i0}
=
\frac{2\pi}{i}\delta(x)
\,.
\end{eqnarray}
In view of the orthogonality of the Dirac functions
and the asymmetry of the functions
$\Psi_{\ground}$ and $\bar{\Psi}_{\ground}$
the first term of
Eq.~(\ref{polerevers})
yields exactly
Eq.~(\ref{lpaEq3x1old2}).
For $\oee_{n_1}>0$ the second term of
(\ref{polerevers})
does not contribute when inserted in
Eq.~(\ref{lpaEq3x1}).
As it was noticed above, for $\oee_{n_1}<0$
the second term does not develop any singularity at the position
of the resonance and can be disregarded.

\par
Having performed the integration over the time variables
($t_{u_1},t_{u_2},t_{d_1},t_{d_2}$)
we arrive at
\begin{eqnarray}
S_{\ground}^{(2)}
\label{lpaEq3x4}
&=&
(-ie)^2 (2\pi)^2 \int  d^{3}{{\zh r}_{u_1}}d^{3}{{\zh r}_{u_2}}
 \, d^{3}{{\zh r}_{d_1}}\, d^{3}{{\zh r}_{d_2}}
\, d{\omega}_{n_1}\, d{\omega}_{n_2} \,\,
\bar{\Phi}_{\ground}({{\zh r}_{u_1}},{{\zh r}_{u_2}})
\nonumber\\
&&
\times
\delta(\omega_{n_1}+\omega_{n_2}-\mee_{\ground}-\omega')
\nonumber\\
&&
\times
\frac{i}{2\pi}
\sum\limits_{n_1}
\frac{\psi_{n_1}({{\zh r}_{u_1}})\bar{\psi}_{n_1}({{\zh r}_{d_1}})}
{\omega_{n_1}-\oee_{n_1}(1-i0)}
\frac{i}{2\pi}
\sum\limits_{n_2}
\frac{\psi_{n_2}({{\zh r}_{u_2}})\bar{\psi}_{n_2}({{\zh r}_{d_2}})}
{\omega_{n_2}-\oee_{n_2}(1-i0)}
\nonumber\\
&&
\times
\delta(\mee_{\ground}+\omega-\omega_{n_1}-\omega_{n_2})
\,\Phi_{\ground}({{\zh r}_{d_1}},{{\zh r}_{d_2}})
\,,
\end{eqnarray}
where
\begin{eqnarray}
\Phi_{\ground}({\zh r}_1,{\zh r}_2)
&=&
\gamma^{\mu_1}A^{(k,\lambda)}_{\mu_1}({\zh r}_1)
\Psi_{\ground}({\zh r}_1,{\zh r}_2)
\,,
\\
{\bar\Phi}_{\ground}({\zh r}_1,{\zh r}_2)
&=&
{\bar\Psi}_{\ground}({\zh r}_1,{\zh r}_2)
\gamma^{\mu_1}A^{\ast(k',\lambda')}_{\mu_1}({\zh r}_1)
\,,
\\
\mee_{\ground}
&=&
\oee_{1s}+\oee_{1s}
\,.
\end{eqnarray}
Formally the expression
(\ref{lpaEq3x4})
is similar to the one in
Eq.~(\ref{lpaEq3})
which has been derived in the one-electron case.
Taking into account interelectron
interaction corrections to the ground state
the function
$\Phi_{\ground}({\zh r}_1,{\zh r}_2)$
and the energy
$\mee_{\ground}$
will become more complicated,
in particular, the function
$\Phi_{\ground}({\zh r}_1,{\zh r}_2)$
will depend on $\omega_{n_1}$, $\omega_{n_2}$.
Nevertheless,
the form of the expression
(\ref{lpaEq3x4}) would remain unchanged.

\par
Below we will employ the resonance approximation
defining the energy
and width of the level such that they will not depend upon
the features of the particular process of scattering.
Since the function $\Phi_{\ground}({\zh r}_1,{\zh r}_2)$
carries all information about the process of scattering
we can assume it to be arbitrary.

\par
\begin{figure}
\centerline{ \mbox{ \epsfxsize=0.20\textwidth \epsffile{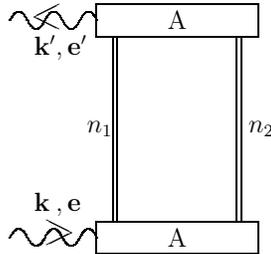} }}
\caption{
The lowest-order amplitude for the photon scattering on
a two-electron ion in the ground state
$\ground$.
In the ground state the interelectron interaction
is taken into account.
}
\label{figure18}
\end{figure}

Accordingly, it is convenient to introduce
a graphical designation:
a rectangle with a letter $\ground$ inside
(see
Fig.~\ref{figure18}).
Lower and upper rectangles represent the functions
$\Phi_{\ground}(x_1,x_2)$ and
${\bar\Phi}_{\ground}(x_1,x_2)$,
respectively,
which are defined as
\begin{eqnarray}
\Phi_{\ground}(x_1,x_2)
&=&
\Phi_{\ground}({\zh r}_1,{\zh r}_2)
e^{-it_1(\mee_{\ground}+\omega)}
\delta(t_1-t_2)
\,,
\\
\bar{\Phi}_{\ground}(x_1,x_2)
&=&
{\bar\Phi}_{\ground}({\zh r}_1,{\zh r}_2)
e^{it_1(\mee_{\ground}+\omega')}
\delta(t_1-t_2)
\,.
\end{eqnarray}
Here
$\Phi_{\ground}({\zh r}_1,{\zh r}_2)$
denotes a complicated vertex function
describing the scattering process
under consideration,
$\mee_{\ground}$
is the energy of the ground state,
$\omega$, $\omega'$
are the frequencies of the absorbed and emitted photons.

\par
We will look for the position of a resonance near
$\omegares=E^{(0)}-\mee_{\ground}+\ordnung(\alpha)$,
where
$E^{(0)}=\oee_{a}+\oee_{b}$
is the energy of two non-interacting electrons.
Applying the identity
(\ref{polerevers})
to
Eq.~(\ref{lpaEq3x4})
one can see that the delta-function term 
is singular close to the resonance, while
the other term remains regular (here we assume the function
$\Phi_{\ground}({\zh r}_1,{\zh r}_2)$ to be arbitrary).
The resonance approximation implies the neglect of
the non-singular (at the resonance) terms in comparison
with singular ones.
Accordingly, within the framework of the resonance approximation
the expression for the S-matrix becomes
\begin{eqnarray}
S_{\ground}^{(2)}
\label{lpaEq3x4old}
&=&
(-ie)^2 (2\pi)^2 \int  d^{3}{{\zh r}_{u_1}}d^{3}{{\zh r}_{u_2}}
 \, d^{3}{{\zh r}_{d_1}}\, d^{3}{{\zh r}_{d_2}}
\, d{\omega}_{n_1}\, d{\omega}_{n_2} \,\,
\bar{\Phi}_{\ground}({{\zh r}_{u_1}},{{\zh r}_{u_2}})
\nonumber\\
&&
\times
\delta(\omega_{n_1}+\omega_{n_2}-\mee_{\ground}-\omega')
\delta(\omega_{n_1}-\oee_{n_1})
\nonumber\\
&&
\times
\frac{i}{2\pi}
\sum\limits_{n_1, n_2}
\frac{\psi_{n_1}({{\zh r}_{u_1}})\bar{\psi}_{n_1}({{\zh r}_{d_1}})
\psi_{n_2}({{\zh r}_{u_2}})\bar{\psi}_{n_2}({{\zh r}_{d_2}})}
{\omega_{n_2}-\oee_{n_2}(1-i0)}
\nonumber\\
&&
\times
\delta(\mee_{\ground}+\omega-\omega_{n_1}-\omega_{n_2})
\Phi_{\ground}({{\zh r}_{d_1}},{{\zh r}_{d_2}})
\,.
\end{eqnarray}
Integrating over
$\omega_{n_1}$, $\omega_{n_2}$
in
Eq.~(\ref{lpaEq3x4old})
and introducing the notation
\begin{eqnarray}
T_{n_1n_2\ground}
&=&
(-e) \int   d^{3}{{\zh r}_1}d^{3}{{\zh r}_2}  \,\,
\bar{\psi}_{n_1}({{\zh r}_1})
\bar{\psi}_{n_2}({{\zh r}_2})
\Phi_{\ground}({{\zh r}_1},{{\zh r}_2})
\,,
\end{eqnarray}
we can express the corresponding amplitude for the scattering
process in a form, similar to
Eq.~(\ref{rsco2}):
\begin{eqnarray}
U_{\ground}^{(2)}
\label{amplituda2}
&=&
\sum_{n_1n_2}
\frac{
  T^{\ast}_{\ground n_1n_2}
  T_{n_1n_2 \ground}
  }{\omega-\oee_{n_1}-\oee_{n_2} + \mee_{\ground}}
\,.
\end{eqnarray}
Since we are searching for the position of the resonance near
$\omegares=E^{(0)}-\mee_{\ground}+\ordnung(\alpha)$,
we have to retain only terms in the denominator of
(\ref{amplituda2})
for which $\oee_{n_1}+\oee_{n_2}=E^{(0)}$.
We assume the energy level close to
$E^{(0)}=\oee_{a}+\oee_{b}$
to be non-degenerate and hence, within the resonance approximation
the amplitude takes the form
\begin{eqnarray}
U_{\ground}^{(2)}
\label{rsco2x031212x2}
&=&
\sum\limits_{n_1 n_2}
\frac{
  T^{\ast}_{\ground n_1n_2}
  T_{n_1n_2 \ground}
  }{\omega-\mee^{(0)} + \mee_{\ground}}
=
T^{+}D^{-1}T
\,.
\end{eqnarray}
Here the summations run only over quantum numbers $n_1, n_2$,
satisfying the condition
$\oee_{n_1}+\oee_{n_2}=E^{(0)}$.
The matrices $T$ and $D^{-1}$ are given by
\begin{eqnarray}
(T)_{n_1 n_2}
&=&
T_{n_1n_2 \ground}
\,,
\\
(D)_{n_1 n_2}
\label{defDmatrixfor2}
&=&
\omega-V^{(0)} + \mee_{\ground}
\end{eqnarray}
together with
\begin{eqnarray}
V^{(0)}
\label{defDmatrixfor56982}
&=&
E^{(0)}
\,.
\end{eqnarray}
As in the one-electron case $T$
defines the type of scattering process under consideration.

\par
\begin{figure}
\centerline{ \mbox{ \epsfxsize=0.10\textwidth \epsffile{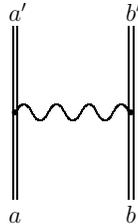} }}
\caption[]{
Feynman graph, describing the first-order interelectron
interaction.
The double solid lines correspond to bound electrons in the field
of the nucleus, the wavy line corresponds to the exchange of virtual
Coulomb and Breit (transverse) photons.
For
$a'=a$
and
$b'=b$
the graph is called ``direct'', and for
$a'=b$,
$b'=a$
it is called ``exchange'' graph, respectively.
}
\label{figure01}
\end{figure}
\begin{figure}
\centerline{ \mbox{ \epsfxsize=0.25\textwidth \epsffile{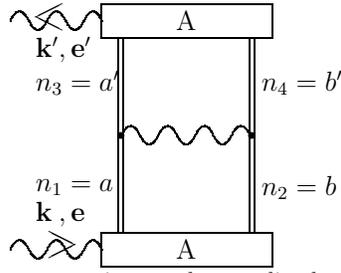} }}
\caption{
First-order of interelectron interaction
 correction to
the amplitude of the photon scattering on a two-electron ion
within the resonance approximation.
}
\label{figure19}
\end{figure}

The interelectron interaction correction
in first order is represented by the graph in
Fig.~\ref{figure01}.
In order to evaluate this contribution 
which also shifts the position the resonance,
one has to consider the graph in
Fig.~\ref{figure19}.
In this paper we employ the Coulomb gauge together with
the covariant metric.
The photon propagator can be written as
\begin{eqnarray}
D_{\mu_1 \mu_2}^{\rm c,t}(x_1,x_2)
\label{defDcmumu}
&=&
\frac{i}{2\pi}
\int\nolimits_{-\infty}^{\infty}d\Omega\,
I_{\mu_1 \mu_2}^{\rm c,t}(|\Omega|,r_{12})\,e^{-i\Omega(t_1-t_2)}
\,,
\end{eqnarray}
where
$r_{12}=|{\zh r}_1-{\zh r}_2|$
and
\begin{eqnarray}
I_{\mu_1 \mu_2}^{\rm c}(\Omega,r_{12})
\label{ic}
&=&
\frac{\delta_{\mu_1 0} \delta_{\mu_2 0}}{r_{12}}
\,,
\\
\label{it}
\label{ib}
I_{\mu_1 \mu_2}^{\rm t}(\Omega,r_{12})
&=&
-
\left(\frac{\delta_{\mu_1 \mu_2}}{r_{12}}\, e^{i\Omega r_{12}}
+
\frac{\partial}{\partial x_1^{\mu_1}}
\frac{\partial}{\partial x_2^{\mu_2}}
\frac{1}{r_{12}}\,
\frac{1-e^{i\Omega r_{12}}}{\Omega^2}
\right)
(1-\delta_{\mu_1 0})(1-\delta_{\mu_2 0})
\,.
\end{eqnarray}
The propagator
$D_{\mu_1 \mu_2}^{\rm c}(x_1,x_2)$
corresponds to Coulomb photons, while
$D_{\mu_1 \mu_2}^{\rm t}(x_1,x_2)$
describes transverse (Breit) photons.
The neglect of retardation implies the substitution
$I_{\mu_1 \mu_2}^{\rm t}(\Omega,r_{12})
=
I_{\mu_1 \mu_2}^{\rm t}(0,r_{12})$.
We employ also the following notation
\begin{eqnarray}
I^{\text{c,t}}(\Omega)_{a'b'ab}
&=&
\label{defmat4}
\sum_{\mu_1\mu_2}
\int d^3{\zh r}_1 d^3{\zh r}_2\,\,
I^{\text{c,t}}_{\mu_1 \mu_2}(\Omega,r_{12})
\left\langle
\bar{\psi}_{a'}({\zh r}_1)
\gamma^{\mu_1}
\psi_{a}({\zh r}_1)
\right\rangle
\left\langle
\bar{\psi}_{b'}({\zh r}_2)
\gamma^{\mu_2}
\psi_{b}({\zh r}_2)
\right\rangle
\,.
\end{eqnarray}
The Lorentz indices $\mu_i$ should indicate that 
the Dirac matrices $\gamma^{\mu_i}$ act on Dirac wave functions
depending on variables ${\zh r}_i$.
The corresponding S-matrix element reads
\begin{eqnarray}
S_{\ground}^{(4)}
\label{lpaEq3x1V1}
&=&
(-ie)^4  \int
d^4 x_1 \, d^4 x_2
\,d\Omega
\,d^{4}{x}_{u_1}\,d^{4}{x}_{u_2}\,d^{4}{x}_{d_1}\,d^{4}{x}_{d_2}
\,d{\omega}_{u_1}\,d{\omega}_{u_2}\,d{\omega}_{d_1}\,d{\omega}_{d_2}
\nonumber\\
&&
\times
{\bar{\Phi}}_{\ground}({\zh r}_{u_1},{\zh r}_{u_2})
e^{it_{u_1}(\mee_{\ground}+\omega')}
\delta(t_{u_1}-t_{u_2})
\nonumber\\
&&
\times
\frac{i}{2\pi}
\sum\limits_{u_1}
\frac{\psi_{u_1}({\zh r}_{u_1})\bar{\psi}_{u_1}({\zh r}_{1})}
{\omega_{u_1}-\oee_{u_1}(1-i0)}
\frac{i}{2\pi}
\sum\limits_{u_2}
\frac{\psi_{u_2}({\zh r}_{u_2})\bar{\psi}_{u_2}({\zh r}_{2})}
{\omega_{u_2}-\oee_{u_2}(1-i0)}
\nonumber\\
&&
\times
e^{-i\omega_{u_1}(t_{u_1}-t_1)}
e^{-i\omega_{u_2}(t_{u_2}-t_2)}
e^{-i\omega_{d_1}(t_1-t_{d_1})}
e^{-i\omega_{d_2}(t_2-t_{d_2})}
\gamma^{\mu_1}\gamma^{\mu_2}
\nonumber\\
&&
\times
\frac{i}{2\pi}
\sum\limits_{d_1}
\frac{\psi_{d_1}({\zh r}_{1})\bar{\psi}_{d_1}({\zh r}_{d_1})}
{\omega_{d_1}-\oee_{d_1}(1-i0)}
\frac{i}{2\pi}
\sum\limits_{d_2}
\frac{\psi_{d_2}({\zh r}_{2})\bar{\psi}_{d_2}({\zh r}_{d_2})}
{\omega_{d_2}-\oee_{d_2}(1-i0)}
\nonumber\\
&&
\times
\frac{i}{2\pi} I_{\mu_1\mu_2}(|\Omega|,r_{12})
e^{-i\Omega(t_1-t_2)}
\nonumber\\
&&
\times
e^{-it_{d_1}(\mee_{\ground}+\omega)}
\delta(t_{d_1}-t_{d_2})
{{\Phi}}_{\ground}({\zh r}_{d_1},{\zh r}_{d_2})
\,,
\end{eqnarray}
while the summations over
$u_1$,$u_2$,$d_1$,$d_2$
run over the entire Dirac spectrum.
Employing the identity
(\ref{polerevers})
and retaining only terms which are singular
near the resonance, i.e. keeping delta-function terms  in
(\ref{polerevers})
only, we obtain
\begin{eqnarray}
S_{\ground}^{(4)}
&=&
-2\pi i \delta(\omega-\omega')
U^{(4)}
\nonumber\\
\label{lpaEq3x1V1x2}
&=&
-2\pi i \delta(\omega-\omega')
\nonumber\\
&&
\times
\sum\limits_{u_1u_2d_1d_2}
T^{+}_{\ground u_1 u_2}
\frac{1}{\mee_{\ground}+\omega-\oee_{u_1}-\oee_{u_2}}
\nonumber\\
&&
\times
e^2 I(|-\oee_{d_1}+\oee_{u_1}|)_{u_1u_2d_1d_2}
\nonumber\\
&&
\times
\frac{1}{\mee_{\ground}+\omega-\oee_{d_1}-\oee_{d_2}}
T_{d_1 d_2 \ground}
\,.
\end{eqnarray}
Within the resonance approximation we are left with terms
satisfying the condition
$\oee_{u_1}+\oee_{u_2}=\oee_{d_1}+\oee_{d_2}=\mee^{(0)}$.

\par
\begin{figure}
\centerline{ \mbox{ \epsfxsize=0.35\textwidth \epsffile{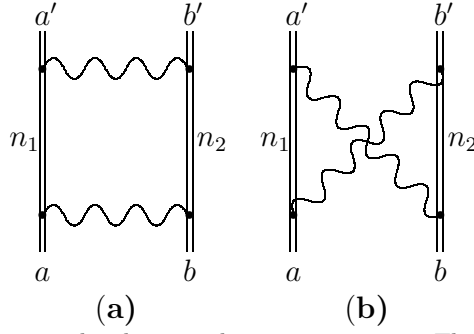} }}
\caption[]{
Feynman graphs describing the second-order interelectron interaction.
The graph (a) is called ``box'' and the graph (b)
is called ``cross''.
Notations are the same as in
Fig.~\ref{figure01}.
The summation over intermediate states is indicated
by $n_1, n_2$.
}
\label{figure02}
\end{figure}
\begin{figure}
\centerline{ \mbox{ \epsfxsize=0.45\textwidth \epsffile{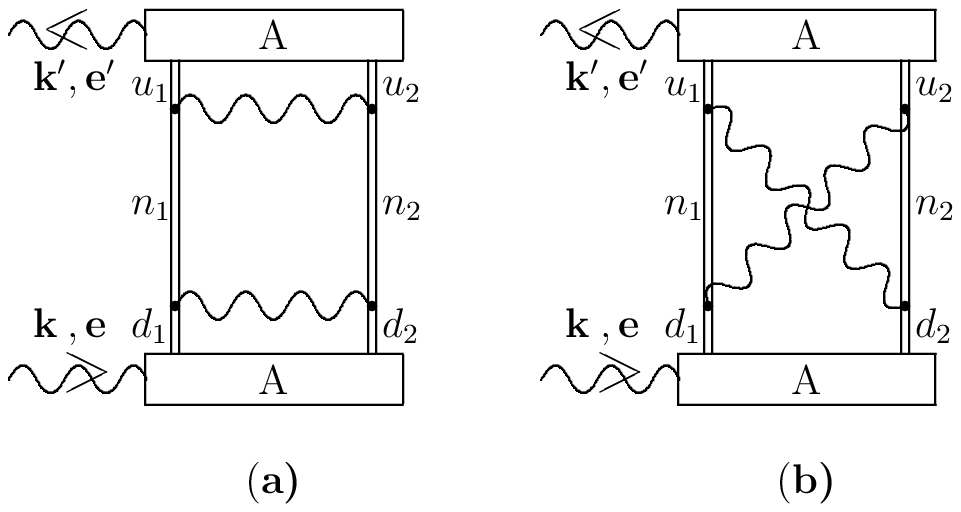} }}
\caption{
Second-order of interelectron interaction correction to
the amplitude of the photon scattering on a two-electron ion
within the resonance approximation.
Graph ({a}) represents the contribution of
the ``box'' graph to the scattering amplitude, and
graph ({b}) the contribution of the ``cross'' graph, respectively
.
}
\label{figure23}
\end{figure}

The second-order interelectron interaction correction is
represented by the graphs in
Fig.~\ref{figure02}.
In order to apply the line profile approach to
the contribution of the ``box'' graph of
Fig.~\ref{figure02}
we have to consider the graph depicted in
Fig.~\ref{figure23}~(a).
The corresponding S-matrix element reads
\begin{eqnarray}
S_{\ground}^{(6)}
\label{lpaEq3x1V186}
&=&
(-ie)^6  \int
d^4 x_1 \, d^4 x_2 \, d^4 x_3 \, d^4 x_4
\,d\Omega \,d\Xi
\,d^{4}{x}_{u_1}\,d^{4}{x}_{u_2}
\,d^{4}{x}_{d_1}\,d^{4}{x}_{d_2}
\,d{\omega}_{u_1}\,d{\omega}_{u_2}
\,d{\omega}_{d_1}\,d{\omega}_{d_2}
\nonumber\\
&&
\times
{\bar{\Phi}}_{\ground}({\zh r}_{u_1},{\zh r}_{u_2})
e^{it_{u_1}(\mee_{\ground}+\omega')}
\delta(t_{u_1}-t_{u_2})
\nonumber\\
&&
\times
\frac{i}{2\pi}
\sum\limits_{u_1}
\frac{\psi_{u_1}({\zh r}_{u_1})\bar{\psi}_{u_1}({\zh r}_{3})}
{\omega_{u_1}-\oee_{u_1}(1-i0)}
\frac{i}{2\pi}
\sum\limits_{u_2}
\frac{\psi_{u_2}({\zh r}_{u_2})\bar{\psi}_{u_2}({\zh r}_{4})}
{\omega_{u_2}-\oee_{u_2}(1-i0)}
\nonumber\\
&&
\times
e^{-i\omega_{u_1}(t_{u_1}-t_3)}
e^{-i\omega_{u_2}(t_{u_2}-t_4)}
e^{-i\omega_{n_1}(t_3-t_{1})}
e^{-i\omega_{n_2}(t_4-t_{2})}
\gamma^{\mu_3}\gamma^{\mu_4}
\nonumber\\
&&
\times
\frac{i}{2\pi}
\sum\limits_{n_1}
\frac{\psi_{n_1}({\zh r}_{3})\bar{\psi}_{n_1}({\zh r}_{1})}
{\omega_{n_1}-\oee_{n_1}(1-i0)}
\frac{i}{2\pi}
\sum\limits_{n_2}
\frac{\psi_{n_2}({\zh r}_{4})\bar{\psi}_{n_2}({\zh r}_{2})}
{\omega_{n_2}-\oee_{n_2}(1-i0)}
\nonumber\\
&&
\times
e^{-i\omega_{d_1}(t_1-t_{d_1})}
e^{-i\omega_{d_2}(t_2-t_{d_2})}
\gamma^{\mu_1}\gamma^{\mu_2}
\nonumber\\
&&
\times
\frac{i}{2\pi}
\sum\limits_{d_1}
\frac{\psi_{d_1}({\zh r}_{1})\bar{\psi}_{d_1}({\zh r}_{d_1})}
{\omega_{d_1}-\oee_{d_1}(1-i0)}
\frac{i}{2\pi}
\sum\limits_{d_2}
\frac{\psi_{d_2}({\zh r}_{2})\bar{\psi}_{d_2}({\zh r}_{d_2})}
{\omega_{d_2}-\oee_{d_2}(1-i0)}
\nonumber\\
&&
\times
\frac{i}{2\pi} I_{\mu_1\mu_2}(|\Xi|,r_{12})e^{-i\Xi(t_1-t_2)}
\frac{i}{2\pi} I_{\mu_3\mu_4}(|\Omega|,r_{34})e^{-i\Omega(t_3-t_4)}
\nonumber\\
&&
\times
e^{-it_{d_1}(\mee_{\ground}+\omega)}
\delta(t_{d_1}-t_{d_2})
{{\Phi}}_{\ground}({\zh r}_{d_1},{\zh r}_{d_2})
\,,
\end{eqnarray}
Employing the identity
(\ref{polerevers})
and retaining only the terms
in the summation over $u_1$,$u_2$ and $d_1$,$d_2$
which become singular close to the resonance,
we obtain the following expression for the S-matrix element
corresponding to the graph in
Fig.~\ref{figure02}~(a)
\begin{eqnarray}
S_{\ground}^{(6)}
&=&
-2\pi i \delta(\omega-\omega')
U^{(6)}
\nonumber\\
\label{lpaopkhgf6}
&=&
-2\pi i \delta(\omega-\omega')
\nonumber\\
&&
\times
\sum\limits_{u_1u_2d_1d_2}
T^{+}_{\ground u_1 u_2}
\frac{1}{\mee_{\ground}+\omega-\oee_{u_1}-\oee_{u_2}}
\nonumber\\
&&
\times
e^4
\frac{i}{2\pi}
\sum\limits_{n_1n_2}
\int d\Omega
\frac{I(|\Omega|)_{u_1u_2n_1n_2}
I(|-\Omega-\oee_{d_1}+\oee_{u_1}|)_{n_1n_2d_1d_2}}
{(-\Omega+\oee_{u_1}-\oee_{n_1}(1-i0))
(\mee_{\ground}+\omega+\Omega-\oee_{u_1}-\oee_{n_2}(1-i0))}
\nonumber\\
&&
\times
\frac{1}{\mee_{\ground}+\omega-\oee_{d_1}-\oee_{d_2}}
T_{ d_1 d_2 \ground}
\,.
\end{eqnarray}
Again within the resonance approximation only terms 
satisfying the condition
$\oee_{u_1}+\oee_{u_2}=\oee_{d_1}+\oee_{d_2}=\mee^{(0)}$
will be kept in the summations over
${u_1}$,${u_2}$,${d_1}$,${d_2}$.

\par
Let us consider separately the reference states terms,
i.e., for which
$\oee_{n_1}+\oee_{n_2}=\mee^{(0)}$
holds.
Inserting a similar identity for the energy denominators
\begin{eqnarray}
\frac{1}
{(-\Omega+\oee_{u_1}-\oee_{n_1}(1-i0))
(\mee_{\ground}+\omega+\Omega-\oee_{u_1}-\oee_{n_2}(1-i0))}
\label{lpa0305110953x}
=
\phantom{1234567890}
\nonumber\\
=
\frac{2\pi}{i}
\frac{\delta(\Omega-\oee_{u_1}+\oee_{n_1})}
{(\mee_{\ground}+\omega-\oee_{n_1}-\oee_{n_2})}
\phantom{123456789012345678901234567890}
\nonumber\\
+
\frac{-1}
{(\Omega-\oee_{u_1}+\oee_{n_1}+\oee_{n_1}i0)
(\mee_{\ground}+\omega+\Omega-\oee_{u_1}-\oee_{n_2}(1-i0))}
\end{eqnarray}
into
Eq.~(\ref{lpaopkhgf6}),
one can verify that the term with the delta-function
coincides with the second element of the geometric progression
for the graph in
Fig.~\ref{figure01}.
Hence, while generating the geometric progression this term will
refer to the second element of the progression.

\par
\begin{figure}
\centerline{ \mbox{ \epsfxsize=0.16\textwidth \epsffile{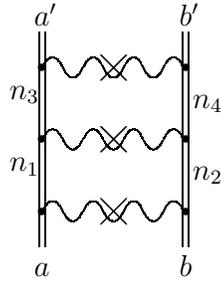} }}
\caption{
The third-order
``box'' Feynman graph.
The notations are the same as in
Fig.~\ref{figure01}.
Here the wavy lines with the cross represent the
sum of the Coulomb and unretarded Breit
interaction.
}
\label{figure03}
\end{figure}

In order to evaluate rigorously the position of the resonance up
to second order in $\alpha$ we have to consider
all corrections of first and second order
simultaneously.
Up to first-order of perturbation theory 
we have to account for 
self-energy (SE) and vacuum-polarization (VP) corrections
as well as for the
exchange of one Coulomb or one Breit photon.
In second order we have to account for
all one- and two-electron Feynman graphs of second order
including radiative corrections,
screening of the self energy and vacuum polarization
and two-photon exchange graphs.
However, the evaluation of the radiative corrections
is not the goal of the present work.
Below we will present the derivation of a formula for
the one-photon exchange contribution in
Fig.~\ref{figure01},
the two-photon exchange ``box'' graph on
Fig.~\ref{figure02}~(a)
and
the three-photon exchange ``box'' graph
Fig.~\ref{figure03}.
\par
The scattering amplitude can be written as
\begin{eqnarray}
\label{twoelprogr5}
U_{\ground a}^{(4)}
&=&
T^{+}D^{-1}\left[(V^{(1)}+V^{(2)}+V^{(3)})D^{-1}\right]T
\,,
\end{eqnarray}
where $V^{(1)}$
corresponds to the one-photon exchange graph
Fig.~\ref{figure01}
(see
Eq.~(\ref{lpaEq3x1V1x2}))
\begin{eqnarray}
V^{(1)}
\label{v1e2ph1x}
&=&
e^2
\sum\limits_{\rm g=c, t}
  I^{\rm g}
  (|-\oee_{a}+\oee_{a'}|)_{a'b'ab}
\,.
\end{eqnarray}
In contrast to the one-electron radiative corrections
(see formula (\ref{v1se})),
this one-photon exchange correction does not depend on
$\omega$.

\par
Taking into account the ``box'' graph in
Fig.~\ref{figure02}~(a)
we obtain (see formulas
(\ref{lpaopkhgf6}), (\ref{lpa0305110953x}))
\begin{eqnarray}
V^{(2)}(\omega)
\label{v22ph}
&=&
e^4\frac{ \,i}{2\pi}
\sum\limits_{\rm g g'=c,t}
\sum\limits_{\oee_{n_1}+\oee_{n_2}\ne \oee_{a}+\oee_{b}}
\int\limits_{-\infty}^{\infty}d\Omega\,
    I^{\rm g}(|\Omega|)_{a' b' n_1n_2}
    I^{\rm g'}(|-\Omega-\oee_{a}+\oee_{a'}|)_{n_1n_2 ab }
\nonumber\\
&&
\times
\frac{1}
{
  (-\Omega + \oee_{a'}- \oee_{n_1}(1-i0))
  (\mee_{\ground}+\omega+\Omega-\oee_{a'}-\oee_{n_2}(1-i0))
}\,
\nonumber\\
&&
+
e^4\frac{\,i}{2\pi}
\sum\limits_{\rm g g'=c,t}
\sum\limits_{\oee_{n_1}+\oee_{n_2}= \oee_{a}+\oee_{b}}
\int\limits_{-\infty}^{\infty}d\Omega\,
    I^{\rm g}(|\Omega|)_{a' b' n_1n_2}
    I^{\rm g'}(|-\Omega-\oee_{a}+\oee_{a'}|)_{n_1n_2 ab}
\nonumber\\
&&
\times
\frac{-1}
{
  (\Omega - \oee_{a'}+\oee_{n_1}+\oee_{n_1}i0)
  (\mee_{\ground}+\omega+\Omega-\oee_{a'}-\oee_{n_2}(1-i0))
}
\,.
\end{eqnarray}
Again the summations over 
${\rm g},{\rm g'}$
run over scalar (Coulomb) and transverse (Breit) photons,
respectively.
The second term in
Eq.~(\ref{v22ph})
represents the remainder after subtracting off 
the reference state singularity.
This subtraction was done at the stage
of generating the geometric progression with
the one-photon exchange insertion.
In particular, the identity
(\ref{lpa0305110953x})
was employed for deriving the expression for
the reference state contribution (the reducible part)
of the ``box'' graph, where the delta-function term
coincides with the second element
of geometric progression for the one-photon exchange graph.

\par
In the case of one-electron ions the reference state term
(see
Eq.~(\ref{v2e1ph2}))
appeared only via the derivative term in
Eq.~(\ref{seseo6nr}).
Considering photon exchange in two-electron ions
the situation is different.
In this case a reference state contribution appears directly in
$V^{(2)}(\omega)$
(see
Eq.~(\ref{v22ph}))
while a derivative term does not arise since 
$V^{(1)}$ does not depend on $\omega$.
Nevertheless, if the sets $\{a,b\}$ and $\{a',b'\}$
are equivalent, the term corresponding to the reference states
can be expressed as a derivative.
At the point of the resonance we can set
$\omega=-\mee_{\ground}+\mee^{(0)}$ in
Eq.~(\ref{v22ph}).
Accordingly, both factors in the denominator will be identical.
Utilizing the formula
\begin{eqnarray}
\frac{-1}{(x+i0)^2}
&=&
\frac{d}{dx}
\frac{1}{(x+i0)}
\end{eqnarray}
and integrating by parts
we can shift the derivative to $I(\Omega)$.

\par
Let us now turn to the three-photon exchange correction
neglecting retardation effects, crossed-photon graphs and
the contribution of the negative energy part of the Dirac spectrum.
Within this approximation
the photon propagator does not depend on the frequency
$\Omega$,
which yields
\begin{eqnarray}
V^{(3)}(\omega)
\label{v3e2ph3x}
&=&
\sum\limits_{\rm g g' g''=c,t}
\mathop{{\sum}'}_{n_1 n_2 n_3 n_4}
{I_{a'b'n_3n_4}^{\rm g}I_{n_3n_4n_1n_2}^{\rm g'}
      I_{n_1n_2ab}^{\rm g''}}
\nonumber\\
&&
\times
\frac{1}{(\oee_{n_3}+\oee_{n_4}-\mee_{\ground}-\omega)
      (\oee_{n_1}+\oee_{n_2}-\mee_{\ground}-\omega)}
\,,
\end{eqnarray}
where the prime at the summation symbol indicates that the
reference states
($\oee_{n_1}+\oee_{n_2}=\oee_{a}+\oee_{b}$,
$\oee_{n_3}+\oee_{n_4}=\oee_{a}+\oee_{b}$)
are omitted.

\par
Taking together the contributions of
$V^{(1)}$, $V^{(2)}$ and $V^{(3)}$,
we can generate a geometric progression,
where the $l$-th term reads
\begin{eqnarray}
\label{twoelprogr8}
Q_l
&=&
T^{+}D^{-1}\left[(V^{(1)}+V^{(2)}+V^{(3)})D^{-1}\right]^{l} T
\,.
\end{eqnarray}
Performing similar steps as in the one-electron case
we sum up this progression and
derive a condition for the position of the resonance
\begin{eqnarray}
V^{(0)}
+
{\real}\{\,
V^{(1)}(\omegares)
+V^{(2)}(\omegares)
+V^{(3)}(\omegares)
\,\}
-\oee_{\ground}
-\omegares
\label{v3e2ph3}
&=&
0
\,.
\end{eqnarray}
The energy and the width of the level will be equal to
(see Eqs.~(\ref{lpaDefE}), (\ref{lpaDefGamma}))
\begin{eqnarray}
E
&=&
V^{(0)}
+
{\real}\{\,
V^{(1)}(\omegares)
+V^{(2)}(\omegares)
+V^{(3)}(\omegares)
\}
+\ordnung(\alpha^4)
\,,
\\
\Gamma
\label{defGammaWidth2el}
&=&
-2\,\,
{\image}\{\,
V^{(1)}(\omegares)
+V^{(2)}(\omegares)
+V^{(3)}(\omegares)
\}
+\ordnung(\alpha^4)
\,.
\end{eqnarray}
It is important to emphasize that
Eq.~(\ref{defGammaWidth2el})
has the meaning of the correction to the width of the level
only if the full set of Feynman graphs of a given order
is under consideration.
Indeed, the contribution of the graph in
Fig.~\ref{figure01}
cancels completely with
a part of the contribution of the self-energy correction, while
the vacuum polarization gives zero contribution to the width.
Such a cancellation
is an immediate consequence
of the Pauli principle according to which
transitions of electrons into occupied states are prohibited
\cite{labzowsky96b}.

\par
We note that the two-electron graphs to the first order in
$V^{(1)}$
does not depend on
$\omega$.
Hence, the solution of
Eq.~(\ref{v3e2ph3})
together
with
$V^{(1)}$,
$V^{(2)}$,
$V^{(3)}$
given by
Eqs.~
(\ref{v1e2ph1x}), (\ref{v22ph}), (\ref{v3e2ph3x})
yields
\begin{eqnarray}
\omegares
\label{el2final}
&=&
-\mee_{\ground}
+
V^{(0)}
+
{\real}\{\,
V^{(1)}(\mee^{(0)}-\mee_{\ground})
+
V^{(2)}(\mee^{(0)}-\mee_{\ground}) \nonumber
\\
&&
+
V^{(3)}(\mee^{(0)}-\mee_{\ground})
+
V^{(1)}(\mee^{(0)}-\mee_{\ground})
\left[\frac{\partial V^{(2)}(\omega)}{\partial\omega}
\right]{\vphantom{\frac{8}{8}}}_{
\textstyle \omega=\mee^{(0)}-\mee_{\ground}}
\,\}
\nonumber\\
&&
+
\ordnung(\alpha^4)
\,.
\end{eqnarray}
The term
$V^{(1)}(\oee_{a}-\oee_{\ground})$
represents the contribution of one-photon exchange graph
Fig.~\ref{figure01},
while the term
$V^{(2)}(\oee_{a}-\oee_{\ground})$
accounts for the contribution of the two-photon exchange
graphs in
Fig.~\ref{figure02}.
In particular, this term includes the contribution of
the reference states occurring in these graphs.
The third-order term
$V^{(3)}(\oee_{a}-\oee_{\ground})$
contributes to the three-photon exchange graphs in
Fig.~\ref{figure03}.
It does not contain the contribution of the reference
states because we
disregard the retardation effects considering it 
within the framework of
relativistic many-body perturbation theory (RMBPT).
The derivative term in
Eq.~(\ref{el2final})
as well as that term in
Eq.~(\ref{seseo6nr})
does not correspond to a certain Feynman graph.
Similar to 
Eq.~(\ref{seseo6nr})
it can be related to the contribution of
the reducible part (reference states) of the graph in
Fig.~\ref{figure03}.

\par
Let us mention that if we would take
into account in addition the radiative corrections and
as well as screening effects,
we would obtain a corresponding correction $V^{(1)}(\omega)$
containing the contribution of the electron self energy
(\ref{v1se}), the one-photon exchange (\ref{v1e2ph1x})
and the vacuum polarization.
For $V^{(2)}(\omega)$
we would similarly obtain the sum of
Eqs.~(\ref{v2e1ph2}), (\ref{v22ph})
and in addition all the missing radiative effects of
second order together with the screened self-energy
and vacuum-polarization corrections.
Accordingly, instead of 
Eq.~(\ref{el2final})
we would have been left with
\begin{eqnarray}
\omegares
\label{finallpa}
&=&
-\mee_{\ground}
+
V^{(0)}
+
{\real}\left\{\,
V^{(1)}(\mee^{(0)}-\mee_{\ground})
+
V^{(2)}(\mee^{(0)}-\mee_{\ground})
  \right.
\nonumber\\
&&
+
V^{(1)}(\mee^{(0)}-\mee_{\ground})
\left[\frac{\partial V^{(1)}(\omega)}{\partial\omega}
\right]
{\vphantom{\frac{8}{8}}}_{\textstyle \omega=\mee^{(0)}-\mee_{\ground}}
+
V^{(3)}(\mee^{(0)}-\mee_{\ground})
\nonumber\\
&&
+
\frac{1}{2}
\,
V^{(1)}(\mee^{(0)}-\mee_{\ground})^2
\left[\frac{\partial^2 V^{(1)}(\omega)}{\partial\omega^2}
\right]{\vphantom{\frac{8}{8}}}_{
\textstyle \omega=\mee^{(0)}-\mee_{\ground}}
\nonumber\\
&&
+
V^{(1)}(\mee^{(0)}-\mee_{\ground})
\left[\frac{\partial V^{(1)}(\omega)}{\partial\omega}
\right]{\vphantom{\frac{8}{8}}}_{\textstyle
\omega=\mee^{(0)}-\mee_{\ground}}^2
\nonumber\\
&&
+
V^{(1)}(\mee^{(0)}-\mee_{\ground})
\left[\frac{\partial V^{(2)}(\omega)}{\partial\omega}
\right]{\vphantom{\frac{8}{8}}}_{
\textstyle \omega=\mee^{(0)}-\mee_{\ground}}
\nonumber\\
&&
\left.
+
V^{(2)}(\mee^{(0)}-\mee_{\ground})
\left[\frac{\partial V^{(1)}(\omega)}{\partial\omega}
\right]{\vphantom{\frac{8}{8}}}_{
\textstyle \omega=\mee^{(0)}-\mee_{\ground}}
\,\right\}
+
\ordnung(\alpha^4)
\,.
\end{eqnarray}

\par
Formulating the line profile approach for $N$-electron ions,
it might be convenient to introduce the function
\begin{eqnarray}
\Phi_{\ground}(x_1,\cdots,x_N)
\label{timedepN}
&=&
\Phi_{\ground}({{\zh r}_1},\cdots,{{\zh r}_N})
e^{-it_1(\mee_{\ground}+\omega)}
\prod\limits^{N}_{j=2}\delta(t_1-t_j)
\,,
\end{eqnarray}
which should be depicted graphically
by a rectangle with a letter $\ground$ inside
and with $N$ out-going electron lines.
Here
$\Phi_{\ground}({{\zh r}_1},\cdots,{{\zh r}_N})$
describes the $N$-electron ions in the lowest (ground) state
$\ground$ together with the absorbed photon.
Accordingly, formula
(\ref{polerevers})
generalizes to
\begin{eqnarray}
\prod\limits^{N}_{j=1}
\frac{1}{(\omega_{n_j}-\oee_{n_j}(1-i0))}
\label{polereversmulti}
&=&
\left[
\prod\limits^{N-1}_{j=1}
\left(
\frac{2\pi}{i}
\delta(\omega_{n_j}-\oee_{n_j})
+
\frac{-1}{(-\omega_{n_j}+\oee_{n_j}+i0\oee_{n_j})}
\right)
\right]
\nonumber\\
&&
\times
\frac{1}{(\omega_{n_N}-\oee_{n_N}(1-i0))}
\,.
\end{eqnarray}
This identity can be written as
\begin{eqnarray}
\prod\limits^{N}_{j=1}
\frac{1}{(\omega_{n_j}-\oee_{n_j}(1-i0))}
\label{polereversmultixf}
&=&
\frac{
\prod\limits^{N-1}_{j=1}
\frac{2\pi}{i}
\delta(\omega_{n_j}-\oee_{n_j})}
{(\omega_{n_N}-\oee_{n_N}(1-i0))}
+
f(\omega,\oee)
\,.
\end{eqnarray}
Then the terms in
Eqs.~(\ref{lpaEq3x4}), (\ref{lpaEq3x1V1}) and (\ref{lpaEq3x1V186}),
which correspond to the function $f(\omega,\oee)$
will not contain the singularities and
will be omitted within the framework of the resonance approximation.
Eqs.~(\ref{el2final}) and (\ref{finallpa})
will remain unchanged,
however, now $V$
will contain additional contributions of
three- and up to $N$-electron graphs.
In particular, for three-electron ions the functions
$V^{(2)}(\omega)$
and
$V^{(3)}(\omega)$
will also account for contributions of three-electron graphs
(see
\cite{andreev01}).

\subsection{Line profile approach for many-electron ions
            (quasi-degenerate energy levels)}
\label{lpa3}
We now turn to the application of the line profile approach
to quasi-degenerate levels.
Without loss of generality, we can restrict ourselves
to two mixing configurations.
We will search for the positions of the resonances corresponding
to these configurations
and will construct basic wave functions
$\Psi_1$
and
$\Psi_2$
within the
\mbox{$j$--$j$} coupling scheme.
The energies corresponding to these wave functions
are denoted by
$\mee^{(0)}_1$,
$\mee^{(0)}_2$
and they
are supposed to be close to the exact energies of
the electron configurations under consideration.
Employing the line profile approach we will consider
a scattering of a photon on a two-electron ion in
its ground state ${\ground}$.
The positions of resonances may be found near the values
$\omegares_1=
\mee^{(0)}_{1}-\mee_{\ground}+\ordnung(\alpha)$
and
$\omegares_2=
\mee^{(0)}_{2}-\mee_{\ground}+\ordnung(\alpha)$,
respectively.
Within the resonance approximation we will have to retain two
terms in the sum
(\ref{amplituda2})
corresponding to the basic functions
$\Psi_1$
and
$\Psi_2$.
The scattering amplitude may be written as
\begin{eqnarray}
\label{twoelprogr61}
U_{\ground a}
&=&
T^{+}D^{-1}\left[\Delta VD^{-1}\right]T
\,,
\end{eqnarray}
where $D$ is a matrix $2\times2$, defined on the functions
$\Psi_1$, $\Psi_2$:
\begin{eqnarray}
\label{twoelprogr62}
D
&=&
\omega+\mee_{\ground}-V^{(0)}
\,,
\\
V^{(0)}
\label{lpaQuasV0}
&=&
\hat{h}_1+\hat{h}_2
\,,
\\
\Delta V
\label{lpaDeltaV}
&=&
V-V^{(0)}
=
V^{(1)}+V^{(2)}+V^{(3)}+\ldots
\,.
\end{eqnarray}
Here $\hat{h}_1$, $\hat{h}_2$
are the one-electron Dirac Hamiltonians
acting on the one-electron Dirac wave functions depending on
${\zh r}_1$ or ${\zh r}_2$, respectively.
Since the functions $\Psi_1$ and $\Psi_2$
are orthogonal the matrix $D$ is diagonal.
Accordingly, we now have to compose a geometric matrix progression
with the $l$-th term
\begin{eqnarray}
\label{twoelprogr9}
Q_l
&=&
T^{+}D^{-1}\left[\Delta VD^{-1}\right]^{l} T
\,
\end{eqnarray}
and sum it up employing the formula for a convergent
geometric progression.
The expression for the amplitude reads
\begin{eqnarray}
U_{\ground }
&=&
T^{+} \left[D-\Delta V\right]^{-1} T
\equiv
T^{+} \frac{1}{D-\Delta V} T
=
T^{+} \frac{1}{\omega+\mee_{\ground}- V} T
\,.
\end{eqnarray}
Introducing the function
${\zh \Phi}=(\Phi_1,\Phi_2)$
by means of the relation
${\zh\Phi}=B{\zh\Psi}$,
where the matrix $B$ is assumed to diagonalize
the matrix \mbox{$V=V^{(0)}+\Delta V$},
i.e. $V^{\text{diag}} = B^{+}VB$.
The expression for the amplitude can now be written
in the following form
\begin{eqnarray}
U_{\ground}
\label{ampl86326}
&=&
T^{+}_{\ground \Phi_1}
\frac{1}{\omega+\mee_{\ground}-[B^{+}VB]_{\Phi_1 \Phi_1}}
T_{\Phi_1 \ground}
+
T^{+}_{\ground \Phi_2}
\frac{1}{\omega+\mee_{\ground}-[B^{+}VB]_{\Phi_2 \Phi_2}}
T_{\Phi_2 \ground}
\nonumber
\\
\label{ampl86326rfg}
&=&
T^{+}_{\ground \Phi_1}
\frac{1}
{\omega+\mee_{\ground}-
V^{\text{diag}}_{\Phi_1 \Phi_1}(\omega)}
T_{\Phi_1 \ground}
+
T^{+}_{\ground \Phi_2}
\frac{1}
{\omega+\mee_{\ground}-
V^{\text{diag}}_{\Phi_2 \Phi_2}(\omega)}
T_{\Phi_2 \ground}
\,.
\end{eqnarray}

\par
Taking a square modulus of the amplitude
(\ref{ampl86326rfg})
and integrating over the directions of the absorbed and
emitted photons yields a line profile for the probability of
photon absorption.
The positions of the resonances are determined by the equations
\begin{eqnarray}
\omegares_1+\mee_{\ground}
-
\real\{V^{\text{diag}}_{\Phi_1 \Phi_1}
(\omegares_1)\}
\label{sdfc1}
&=&0
\,,
\\
\omegares_2+\mee_{\ground}
-
\real\{V^{\text{diag}}_{\Phi_2 \Phi_2}
(\omegares_2)\}
\label{sdfc2}
&=&0
\,.
\end{eqnarray}
Hence, the energy of the configurations are:
\begin{eqnarray}
E_{\Phi_1}
\label{kjdj1}
&=&
\real\{V^{\text{diag}}_{\Phi_1 \Phi_1}
(\omegares_1)\}
\,,
\\
E_{\Phi_2}
\label{kjdj2}
&=&
\real\{V^{\text{diag}}_{\Phi_2 \Phi_2}
(\omegares_2)\}
\,.
\end{eqnarray}
Assuming that the energies of the configurations are
close to each other, we can expand
Eqs.~(\ref{sdfc1}), (\ref{sdfc2})
into a Taylor series around the values
$\omegares_1=-\mee_{\ground}+\mee^{(0)}_1$
and
$\omegares_2=-\mee_{\ground}+\mee^{(0)}_2$,
respectively.
As in the case of non-degenerate levels
this can be achieved up to any desired accuracy.

\par
Note, that employing the resonance approximation
in case of non-degenerate level we have to retain
in a corresponding sum (\ref{amplituda2})
certain many-electron functions
composed within the \mbox{$j$--$j$} coupling scheme.
Indeed, after diagonalization of the matrix $V$
all other combinations of one-electron functions will yield zero
in view of the antisymmetry of the wave function of the ground
state and the symmetry of the matrix $V$.
Hence, having constructed a many-electron function in
the \mbox{$j$--$j$} coupling scheme,
Eq.~(\ref{twoelprogr5})
becomes a scalar one.

\par
The line profile approach outlined above can be easily employed
for an arbitrary number of degenerate levels.
The generalization of the method to $N$-electron ions
was described at the end of the previous section.

\section{Evaluation of the energy levels of
quasi-degenerate two-electron configurations}
\label{evaluation}
We will evaluate the interelectron interaction correction
for the two-electron configurations
$(1s2p)2 {}^1\! P_{1}$ and $(1s2p)2 {}^3\! P_{1}$.
Employing the relativistic \mbox{$j$--$j$} coupling scheme
these energy levels become quasi-degenerate in the region
$Z\le 40$.
To treat such states within the framework of QED
we will apply the line profile approach.
Within the \mbox{$j$--$j$} coupling scheme
the wave function of a two-electron configuration
can be represented as
\begin{eqnarray}
\Psi_{JMj_1j_2l_1l_2}({\zh r}_{1}, {\zh r}_{2})
\label{wavef}
\label{funcxjj}
&=&
N
\sum\limits_{m_1 , m_2}
\mbox{\rm C}^{j_1j_2}_{JM}(m_1m_2)
\nonumber\\
&&
\times
\left[
\psi_{j_1l_1m_1}({\zh r}_{1})
\psi_{j_2l_2m_2}({\zh r}_{2})
-
\psi_{j_1l_1m_1}({\zh r}_{2})
\psi_{j_2l_2m_2}({\zh r}_{1})
\right]
\,,
\end{eqnarray}
where the normalization constant is $N=1/2$
for equivalent electrons and $N=1/\sqrt{2}$
for non-equivalent electrons, respectively.
$\mbox{\rm C}^{j_1j_2}_{JM}(m_1m_2)$ is a Clebsch-Gordan
coefficient.
The one-electron Dirac functions
$\psi_{jlm}({\zh r})$
are characterized by the standard set of one-electron
quantum numbers: total angular momentum $j$,
its projection $m$ and the orbital angular momentum $l$,
that fixes the parity of the state.
For the two-electron wave function the relevant quantum numbers
are the total angular momentum $J$ and its projection $M$.

\par
Following the procedure described in
Section~\ref{lpa3}
we will construct the matrix $V$
(\ref{lpaDeltaV})
on the functions
(\ref{funcxjj})
\begin{eqnarray}
\Psi_{J=1,M=0,j_1=1/2,j_2=1/2,l_1=0,l_2=1}
&\equiv&
(1s2p_{1/2})
\,,
\\
\Psi_{J=1,M=0,j_1=1/2,j_2=3/2,l_1=0,l_2=1}
&\equiv&
(1s2p_{3/2})
\,
\end{eqnarray}
and examine the positions of the resonances close to
$\omegares_1
=-\mee_{A}+\oee_{1s}+\oee_{2p_{1/2}}+\ordnung(\alpha)$
and
$\omegares_2
=-\mee_{A}+\oee_{1s}+\oee_{2p_{3/2}}+\ordnung(\alpha)$,
respectively.

\par
As it has been elaborated in
Section~\ref{lpa}
the operator $V$, in general, depends on $\omega$.
The position of the resonance can be derived via Taylor
expansion at the approximate positions of the resonances
$\omegares_1=-\mee_{A}+\oee_{1s}+\oee_{2p_{1/2}}$
and
$\omegares_2=-\mee_{A}+\oee_{1s}+\oee_{2p_{3/2}}$
(see
Eq.~(\ref{el2final})).
For the practical calculations it is convenient to
expand some matrix elements of $V$ at the point $\omegares_1$
and others at the point $\omegares_2$
keeping only terms $\ordnung(\alpha^2)$ in
both expansions.
The resulting inaccuracy can be referred to corrections
$\ordnung(\alpha^3)$
\cite{shabaev02},
because at low $Z$ values the energy difference
$\oee_{2p_{1/2}}-\oee_{2p_{3/2}}$ becomes small,
while at large-$Z$ values the degeneracy of the levels
$2 {}^1\! P_{1}$, $2 {}^3\! P_{1}$
is nearly negligible.

\par
The interelectron interaction correction is represented
by the set of graphs
(Figs.~\ref{figure01} and \ref{figure02})
which is symmetric under interchange of
the upper and the lower indices and
relabeling of the electrons in the graphs.
Accordingly, the operator $V$ is given by a symmetric
(and in general complex) matrix.
However, as a consequence of performing the Taylor expansion of
the matrix elements of $V$
and neglecting third- and higher-orders terms
it can lead to
a non-symmetrical matrix.
To prevent this asymmetry arising due to purely technical
reasons one may symmetrize the matrix $V$ by hand.

\par
Hence, the matrix elements of the frequency-dependent
operator $V$
evaluated at the resonances 
can be written as
\begin{eqnarray}
\langle(1s2p_{1/2})|V^{}(\omegares)|(1s2p_{1/2})\rangle
\label{resquas1}
&=&
\langle(1s2p_{1/2})|F^{}|(1s2p_{1/2})\rangle
\,,
\\
\langle(1s2p_{3/2})|V^{}(\omegares)|(1s2p_{3/2})\rangle
&=&
\langle(1s2p_{3/2})|F^{}|(1s2p_{3/2})\rangle
\,,
\\
\langle(1s2p_{1/2})|V^{}(\omegares)|(1s2p_{3/2})\rangle
&=&
\frac{1}{2}
\left[\langle(1s2p_{1/2})|F^{}|(1s2p_{3/2})\rangle
\vphantom{F^{(1)}}\right.
\nonumber\\
&&
+
\left.\vphantom{F^{(1)}}
\langle(1s2p_{3/2})|F^{}|(1s2p_{1/2})\rangle\right]
\,,
\\
\langle(1s2p_{3/2})|V^{}(\omegares)|(1s2p_{1/2})\rangle
&=&
\langle(1s2p_{1/2})|V^{}(\omegares)|(1s2p_{3/2})\rangle
\,.
\end{eqnarray}
The operator $F$ is defined via its action on
the set of the one-electron Dirac functions $\{ab\}$, which
in our case consists of
$\{ab\}= \{ 1s2p_{1/2} \} ,\{ 1s2p_{3/2} \}$.
To zeroth-order perturbation theory the operator
$F$ reads
(see Eq.~(\ref{lpaQuasV0})):
\begin{eqnarray}
F^{(0)}_{a'b'ab}
&=&
\oee_{a}\delta_{{a'},{a}}
+\oee_{b}\delta_{{b'},{b}}
\,.
\end{eqnarray}
Being interested in ionization energies it is more convenient
to introduced a shifted $F$ with the zeroth-order matrix element
\begin{eqnarray}
F^{(0)}_{a'b'ab}
\label{DefUsedBoundEnergy}
&=&
\oee_{a}\delta_{{a'},{a}}
+\oee_{b}\delta_{{b'},{b}}
-\oee_{1s}-m
\,,
\end{eqnarray}
where the electron rest energy $m$
(in relativistic unites)
and ${1s}$-electron energy are subtracted.
In first-order perturbation theory
the interelectron interaction represented by the graph in
Fig.~\ref{figure01}
can be described by the matrix element
\begin{eqnarray}
F_{a'b'ab}^{(1)}
\label{qboxirrFant}
&=&
e^2
I(|\oee_{a'}-\oee_{a}|)_{a'b'ab}
\,.
\end{eqnarray}
Since graph in
Fig.~\ref{figure01}
is irreducible $F_{a'b'ab}^{(1)}$ coincides with
the expression
(Eq.~(\ref{v1e2ph1x}))
for non-degenerate levels.
In second-order perturbation theory
we have to account for the two-photon exchange
corrections depicted in
Fig.~\ref{figure02}:
\begin{eqnarray}
F_{a'b'ab}^{(2)(\rm box,irr)}
\label{qboxirr}
&=&
e^4
\mathop{{\sum}}_{\rm g g'}
\mathop{{\sum}}_{n_1 n_2}
(1-\delta_{\mee^{(0)}_{n_1n_2},\mee^{(0)}_{ab}})
\nonumber\\
&&
\times
\left\{ \vphantom{\frac{8}{8}}\right.
\frac{i}{2\pi}\int\limits_{-\infty}^{\infty}d\Omega\,
\frac{I^{\rm g}(|\Omega|)_{a'b'n_1n_2}
I^{\rm g'}(|\Omega-\oee_{a'}+\oee_a|)_{n_1n_2ab}
}
{(\mee^{(0)}_{ab}-\mee^{(0)}_{n_1n_2})
 (\Omega-\oee_{n_2}+\mee^{(0)}_{ab}-\oee_{a'}+i0\oee_{n_2})}
\nonumber\\
&&
+
\frac{i}{2\pi}\int\limits_{-\infty}^{\infty}d\Omega\,
\frac{I^{\rm g}(|\Omega|)_{b'a'n_1n_2}
I^{\rm g'}(|\Omega-\oee_{a}+\oee_{a'}|)_{n_1n_2ba}
}
{(\mee^{(0)}_{ab}-\mee^{(0)}_{n_1n_2})
 (\Omega-\oee_{n_2}+\oee_{a'}+i0\oee_{n_2})}
\left.\vphantom{\frac{8}{8}}\right\}
\,,
\end{eqnarray}
\begin{eqnarray}
F_{a'b'ab}^{(2)(\rm box,red)}
\label{qboxred}
&=&
-\frac{1}{2}
e^4
\mathop{{\sum}}_{\rm g g'}
\mathop{{\sum}}_{n_1 n_2}
\nonumber\\
&&
\times
\left\{ \vphantom{\frac{8}{8}}\right.
\delta_{\mee^{(0)}_{n_1n_2},\mee^{(0)}_{ab}}
\left[\vphantom{\frac{8}{8}}\right.
\frac{i}{2\pi}\int\limits_{-\infty}^{\infty}d\Omega\,
\frac{I^{\rm g}(|\Omega|)_{a'b'n_1n_2}
I^{\rm g'}(|\Omega-\oee_{a'}+\oee_a|)_{n_1n_2ab}
}
{(\Omega-\oee_{n_2}+\mee^{(0)}_{ab}-\oee_{a'}+i0\oee_{n_2})^2}
\nonumber\\
&&
+
\frac{i}{2\pi}\int\limits_{-\infty}^{\infty}d\Omega\,
\frac{I^{\rm g}(|\Omega|)_{b'a'n_1n_2}
I^{\rm g'}(|\Omega-\oee_{a}+\oee_{a'}|)_{n_1n_2ba}
}
{(\Omega-\oee_{n_2}+\oee_{a'}+i0\oee_{n_2})^2}
\left.\vphantom{\frac{8}{8}}\right]
\nonumber\\
&&
+
\delta_{\mee^{(0)}_{n_1n_2},\mee^{(0)}_{a'b'}}
(1-\delta_{\mee^{(0)}_{n_1n_2},\mee^{(0)}_{ab}})
\nonumber\\
&&
\times
\left[\vphantom{\frac{8}{8}}\right.
\frac
{I(|\oee_{n_2}-\mee^{(0)}_{ab}+\oee_{a'}|)_{a'b'n_1n_2}
I(|\oee_{n_2}-\mee^{(0)}_{ab}+\oee_{a}|)_{n_1n_2ab}}
{\mee^{(0)}_{ab}-\mee^{(0)}_{n_1n_2}}
\nonumber\\
&&
+
\frac
{I(|\oee_{n_2}-\oee_{a'}|)_{b'a'n_1n_2}
I(|\oee_{n_2}-\oee_{a}|)_{n_1n_2ba}}
{\mee^{(0)}_{ab}-\mee^{(0)}_{n_1n_2}}
\left.\left. \vphantom{\frac{8}{8}}\right]\right\}
\,,
\end{eqnarray}
\begin{eqnarray}
F_{a'b'ab}^{(2)(\rm cross,irr)}
\label{qcrossirr}
&=&
e^4
\mathop{{\sum}}_{\rm g g'}
\mathop{{\sum}}_{n_1 n_2}
\left\{ \vphantom{\frac{8}{8}}\right.
(1-\delta_{0,(\oee_{n_2}-\oee_{n_1}+\oee_{b}-\oee_{a'})})
\nonumber\\
&&
\times
\frac{i}{2\pi}\int\limits_{-\infty}^{\infty}d\Omega
\frac{I^{\rm g}(|\Omega|)_{b'n_2n_1a}
I^{\rm g'}(|\Omega-\oee_{a'}+\oee_{a}|)_{n_1a'bn_2}
}
{(\oee_{n_2}-\oee_{n_1}+\oee_{b}-\oee_{a'})
 (\Omega-\oee_{n_2}+\oee_{a}+i0\oee_{n_2})}
\nonumber\\
&&
+
(1-\delta_{0,(\oee_{n_2}-\oee_{n_1}-\oee_{b}+\oee_{a'})})
\nonumber\\
&&
\times
\frac{i}{2\pi}\int\limits_{-\infty}^{\infty}
d\Omega\,
\frac{I^{\rm g}(|\Omega|)_{n_1b'an_2}
I^{\rm g'}(|\Omega-\oee_{a'}+\oee_{a}|)_{a'n_2n_1b}
}
{(\oee_{n_2}-\oee_{n_1}-\oee_{b}+\oee_{a'})
 (\Omega-\oee_{n_2}+\mee^{(0)}_{ab}-\oee_{a'}+i0\oee_{n_2})}
\left.\vphantom{\frac{8}{8}}
\right\}
\,,
\end{eqnarray}
\begin{eqnarray}
F_{a'b'ab}^{(2)(\rm cross,red)}
\label{qcrossred}
&=&
e^4
\mathop{{\sum}}_{\rm g g'}
\mathop{{\sum}}_{n_1 n_2}
\delta_{0,(\oee_{n_2}-\oee_{n_1}+\oee_{b}-\oee_{a'})}
\nonumber\\
&&
\times
\frac{i}{2\pi}
\int\limits_{-\infty}^{\infty}d\Omega
\frac{I^{\rm g}(|\Omega|)_{b'n_2n_1a}
I^{\rm g'}(|\Omega-\oee_{a'}+\oee_a|)_{n_1a'bn_2}
}
{(\Omega-\oee_{n_2}+\oee_{a}+i0\oee_{n_2})^2}
\,.
\end{eqnarray}
In
Eqs.~(\ref{qboxirr}) and (\ref{qboxred})
the notations
$\mee^{(0)}_{ab}=\oee_{a}+\oee_{b}$,
$\mee^{(0)}_{a'b'}=\oee_{a'}+\oee_{b'}$,
$\mee^{(0)}_{n_1n_2}=\oee_{n_1}+\oee_{n_2}$
are introduced.
Index ${\rm g}$ runs over ${\rm c,t}$
(scalar and transverse photons).
The Kronecker symbols ensure that terms with potentially
zero denominators will be omitted in the summation over
$n_1,n_2$.

\par
Note , that the
Eqs.~(\ref{qboxirr}), (\ref{qcrossirr}), (\ref{qcrossred})
for the irreducible parts
coincide generically with
Eq.~(\ref{v22ph})
for non-degenerate levels (see also
\cite{andreev03}).
However, for the reducible part of the ``box'' graph additional
terms originating from the geometric progression
for the one-photon exchange  graph
(non-diagonal matrix elements
of the second term of the progression) occur.
It is easy to make sure that the contribution of
reference states
($\mee^{(0)}_{n_1n_2}=\mee^{(0)}_{ab}$,
$\mee^{(0)}_{n_1n_2}=\mee^{(0)}_{a'b'}$)
to the exchange of two Coulomb photons 
(or Breit photons with neglect of retardation)
is absent.

\section{Numerical results and their analysis}
\label{results}
The results of the numerical calculations are presented in
Tables~\ref{txquasidegx1},~\ref{txquasidegx3}.
To account for nuclear size corrections
we solved the Dirac equation with the Coulomb
potential generated by a nuclear charge density
described by a Fermi distribution.
The parameters of the Fermi distribution are taken from
Ref.~\cite{andreev03}.

\par
In
Table~\ref{txquasidegx1}
we present
a detailed analysis of our results obtained for
the photon-exchange contribution.
The value $V^{(0)}$
is the binding energy of the $2p$ state according to
Eq.~(\ref{DefUsedBoundEnergy}),
the value $V^{(1)}$ corresponds to
the one-photon exchange contribution
Eq.~(\ref{qboxirrFant})
and $V^{(2)}$ represents the two-photon exchange contributions
given by
Eqs.~(\ref{qboxirr}--\ref{qcrossred}).
We note, that, in general, the matrix $V$ has complex elements
and both their real and imaginary parts contribute
to the energy eigenvalues,
i.e., the real part of the diagonalized matrix $V$.
In our calculation the imaginary part of $V^{(1)}$
is taken into account, while
the imaginary part of the two-photon exchange contribution
($V^{(2)}$) is neglected.
The values
$\Eexch(2 {}^1\! P_{1})$ and $\Eexch(2 {}^3\! P_{1})$
denote the photon-exchange contribution to
the energies of the corresponding electron configurations
(neglecting the radiative corrections).
For $Z=10,18$ we present also the values for the difference between
the energies of the levels under consideration reported in
Ref.~\cite{lindgren01}.

\par
In order to analyse the influence of
the quasi-degeneracy on QED effects in more detail
we compile the corresponding energy shifts of the levels due to
the photon-exchange contribution calculated
within various approximations.
The differences between the energies of the levels calculated
without the approximations
and the energies calculated within the framework of
the approximations are presented in
Table~\ref{txquasidegx1}
[$\Delta \Eexch(2 {}^{1,3}\! P_{1})$: Appr.~1-5].

\par
Approximation~1:
We omit the non-diagonal elements of the matrix
$V=V^{(0)}+V^{(1)}+V^{(2)}$.
Consequently, effects of the quasi-degeneracy are totally 
neglected.

\par
Approximation~2:
We omit the non-diagonal elements only in the matrix $V^{(2)}$.
As stated above the expression for
the one-photon exchange correction
(\ref{qboxirrFant})
does not depend on $\omega$
and coincides with the one for the non-degenerate case.
Accordingly, the first-order contribution is taken into
account just as the solution of the secular equation,
i.e. following usual techniques developed in quantum mechanics
for treating degenerate levels.
The influence of quasi-degeneracy due to the second-order
matrix element $V^{(2)}$ is neglected.

\par
Approximation~3:
We calculate the matrix  elements of $V$ within
the framework of RMBPT.
Compared with the full ab initio QED calculation the
following contributions are missing:
1) negative-energy intermediate states,
2) crossed-photon interaction,
3) rigorous treatment of retardation effects.
As mentioned above within the framework of RMBPT
no contribution due to reference states
(for two-photon exchange) arises.
Accordingly, the energies of the levels just follow
as solutions of the secular equation.

\par
Approximation~4:
Only the matrix elements of $V^{(2)}$
are evaluated within the framework of RMBPT.
According to the comment made on approximation 2,
this also follows the quantum mechanical
treatment for quasi-degeneracy.

\par
Approximation~5:
We neglect the imaginary part of the elements of the matrix $V$.
The matrix $V$ defined in
Eq.~(\ref{lpaDeltaV})
is a complex one.
Although the energy of the level is defined as the real
part of the diagonalized matrix $V(\omega)$
at the point of the resonance ($\omega=\omegares$),
the imaginary part of the elements of 
the matrix $V$ (non-diagonal)
contributes to the energy.

\par
The results in
Table~\ref{txquasidegx1}
demonstrate that for the 
$2 {}^{1}\! P_{1}$, $2 {}^{3}\! P_{1}$ levels a
complete ab initio QED theory for describing the quasi-degeneracy
has to be employed only when going beyond the level 
of second-order corrections.
For $Z<30$ approximation 3 provides an accuracy of
about 1\%
at the level of second-order perturbation theory.
Accordingly, the inaccuracy can be referred
to corrections of third order.
For $Z<60\sim 70$ approximation 4 leads to an
inaccuracy comparable in magnitude with
the corrections of third order.
For $Z>60\sim 70$ the effect of quasi-degeneracy
decreases definitely to the level of third-order corrections.
Consequently, approximation 2 can be employed for high-$Z$ systems.
For $Z>80$ the quasi-degeneracy becomes completely negligible,
i.e., it will be sufficient to employ approximation 1.
The contribution of imaginary parts of the matrix elements $V$
to the energy levels appears as an effect of quasi-degeneracy,
which originates completely from QED.
It is perceptible only for high $Z>70$,
which also reveals that the neglect of the imaginary
part of $V^{(2)}$ has been legitimate.

\par
In
Table~\ref{txquasidegx3}
we present the data for the total energies of
the $2 {}^{1}\! P_{1}$ and $2 {}^{3}\! P_{1}$
two-electron configurations, respectively.
The numbers present the ionization energy of the $2p$-electron with
the opposite sign.
These data are compared with the results obtained by
Plante {\it et al.}~\cite{plante94}
and
Drake~\cite{drake88}.
Two different approximate methods have been employed in these works:
the ``relativistic all-order theory'' (AO)
\cite{plante94}
and ``the unified theory''
\cite{drake88}.
The latter methods account approximately for QED effects,
such as retardation, crossed-photon graphs
and negative-energy intermediate states,
while taking into account partially higher-orders
of the perturbation theory.
In the present work the photon exchange is taken into account up to
the second order.
The self-energy (SE) and vacuum-polarization (VP) corrections
are included only in first order.
Quantitative results for SE and VP corrections are taken from
Refs.~\cite{mohr92,mohr93,soff88,mohr98}.
The SE and VP screening corrections,
the radiative corrections of the second order
and all the corrections of the third and higher orders are omitted.
We note, that the VP screening corrections for the states considered
have been evaluated by Artemyev {\it et al.}
\cite{artemyev00},
while results for the SE screening corrections are not yet available.
Since the SE and VP screening corrections partially
cancel each other, we do not include the results of
\cite{artemyev00}
in
Table~\ref{txquasidegx3}.
In
Table~\ref{txquasidegx5}
we present various theoretical and experimental data
for $2 {}^3\! P_{1}-2 {}^1\! P_{1}$ transition energies.
We conclude that
the discrepancy between our data and those from other results arising
for small values of $Z$  is caused by third- and higher-orders of
the perturbation theory which
have not been accounted for in the present paper.
For high $Z$ the major inaccuracy is due to missing
self-energy, vacuum-polarization
screening corrections and one-electron radiative corrections
of second order.

\acknowledgments
\label{acknowledgement}
The authors are indebted to Prof.~W.~Nagel from the center
of high-performance computing at
TU Dresden for providing access to all
necessary computer facilities.
O.Y.A. is grateful to TU Dresden for the
hospitality during his visits in 2002 and 2003 and to the DFG for
financial support.
The work of O.Y.A. and L.N.L. was supported by the RFBR Grant
No. 02-02-16578
and by Minobrazovanie grant No. E02-3.1-7.
G.P. and G.S. acknowledge financial support from BMBF, DFG and GSI.
%
%
%
%
%
%

\begin{thebibliography}{10}
\providecommand*{\bibinfo}[2]{#2}
\providecommand*{\eprint}[1]{#1}
\providecommand*{\url}[1]{#1}
\bibitem{drake88}
\bibinfo{author}{G.~W. Drake}, \bibinfo{journal}{Can. J. Phys.}
  \bibinfo{volume}{\textbf{66}}, \bibinfo{pages}{586} (\bibinfo{date}{1988}).
\bibitem{plante94}
\bibinfo{author}{D.~R. Plante}, \bibinfo{author}{W.~R. Johnson}, and
  \bibinfo{author}{J.~Sapirstein}, \bibinfo{journal}{Phys. Rev. A}
  \bibinfo{volume}{\textbf{49}}, \bibinfo{pages}{3519} (\bibinfo{date}{1994}).
\bibitem{mohr98}
\bibinfo{author}{P.~J. Mohr}, \bibinfo{author}{G.~Plunien}, and
  \bibinfo{author}{G.~Soff}, \bibinfo{journal}{Phys. Rep.}
  \bibinfo{volume}{\textbf{293}}, \bibinfo{pages}{227} (\bibinfo{date}{1998}).
\bibitem{yerokhin03}
\bibinfo{author}{V.~A. Yerokhin}, \bibinfo{author}{P.~Indelicato}, and
  \bibinfo{author}{V.~M. Shabaev}, \bibinfo{journal}{Phys. Rev. Lett.}
  \bibinfo{volume}{\textbf{91}}, \bibinfo{pages}{073001}
  (\bibinfo{date}{2003}).
\bibitem{klimchitskaya71}
\bibinfo{author}{{G. L. Klimchitskaya and L. N. Labzowsky, Zh. Eksp. Teor. Fiz.
  {\bf 60}, 2019 (1971) [Engl. Transl. Sov. Phys.- JETP {\bf 33}, 1088
  (1971)].}}
\bibitem{labzowsky93b}
\bibinfo{author}{L.~Labzowsky}, \bibinfo{author}{G.~Klimchitskaya}, and
  \bibinfo{author}{\mbox{Yu}. Dmitriev}, \bibinfo{title}{\emph{Relativistic
  Effects in the Spectra of Atomic Systems}} (\bibinfo{publisher}{Institute of
  Physics Publishing}, Bristol and Philadelphia, \bibinfo{year}{1993}).
\bibitem{blundell93}
\bibinfo{author}{S.~Blundell}, \bibinfo{author}{P.~J. Mohr},
  \bibinfo{author}{W.~R. Johnson}, and \bibinfo{author}{J.~Sapirstein},
  \bibinfo{journal}{Phys. Rev. A} \bibinfo{volume}{\textbf{48}},
  \bibinfo{pages}{2615} (\bibinfo{date}{1993}).
\bibitem{lindgren95}
\bibinfo{author}{I.~Lindgren}, \bibinfo{author}{H.~Persson},
  \bibinfo{author}{S.~Salomonson}, and \bibinfo{author}{L.~Labzowsky},
  \bibinfo{journal}{Phys. Rev. A} \bibinfo{volume}{\textbf{51}},
  \bibinfo{pages}{1167} (\bibinfo{date}{1995}).
\bibitem{yerokhin00}
\bibinfo{author}{V.~A. Yerokhin}, \bibinfo{author}{A.~N. Artemyev},
  \bibinfo{author}{V.~M. Shabaev}, \bibinfo{author}{M.~M. Sysak},
  \bibinfo{author}{O.~M. Zherebtsov}, and \bibinfo{author}{G.~Soff},
  \bibinfo{journal}{Phys. Rev. Lett.} \bibinfo{volume}{\textbf{85}},
  \bibinfo{pages}{4699} (\bibinfo{date}{2000}).
\bibitem{mohr00}
\bibinfo{author}{P.~J. Mohr} and \bibinfo{author}{J.~Sapirstein},
  \bibinfo{journal}{Phys. Rev. A} \bibinfo{volume}{\textbf{62}},
  \bibinfo{pages}{052501} (\bibinfo{date}{2000}).
\bibitem{andreev01}
\bibinfo{author}{O.~\mbox{Yu}. Andreev}, \bibinfo{author}{L.~N. Labzowsky},
  \bibinfo{author}{G.~Plunien}, and \bibinfo{author}{G.~Soff},
  \bibinfo{journal}{Phys. Rev. A} \bibinfo{volume}{\textbf{64}},
  \bibinfo{pages}{042513} (\bibinfo{date}{2001}).
\bibitem{asen02}
\bibinfo{author}{B.~\r{A}s\'en}, \bibinfo{author}{S.~Salomonson}, and
  \bibinfo{author}{I.~Lindgren}, \bibinfo{journal}{Phys. Rev. A}
  \bibinfo{volume}{\textbf{65}}, \bibinfo{pages}{032516}
  (\bibinfo{date}{2002}).
\bibitem{andreev03}
\bibinfo{author}{O.~\mbox{Yu}. Andreev}, \bibinfo{author}{L.~N. Labzowsky},
  \bibinfo{author}{G.~Plunien}, and \bibinfo{author}{G.~Soff},
  \bibinfo{journal}{Phys. Rev. A} \bibinfo{volume}{\textbf{67}},
  \bibinfo{pages}{012503} (\bibinfo{date}{2003}).
\bibitem{indelicato01}
\bibinfo{author}{P.~Indelicato} and \bibinfo{author}{P.~J. Mohr},
  \bibinfo{journal}{Phys. Rev. A} \bibinfo{volume}{\textbf{63}},
  \bibinfo{pages}{052507} (\bibinfo{date}{2001}).
\bibitem{artemyev00}
\bibinfo{author}{A.~N. Artemyev}, \bibinfo{author}{T.~Beier},
  \bibinfo{author}{G.~Plunien}, \bibinfo{author}{V.~M. Shabaev},
  \bibinfo{author}{G.~Soff}, and \bibinfo{author}{V.~A. Yerokhin},
  \bibinfo{journal}{Phys. Rev. A} \bibinfo{volume}{\textbf{62}},
  \bibinfo{pages}{022116} (\bibinfo{date}{2000}).
\bibitem{yerokhin99}
\bibinfo{author}{V.~A. Yerokhin}, \bibinfo{author}{A.~N. Artemyev},
  \bibinfo{author}{T.~Beier}, \bibinfo{author}{G.~Plunien},
  \bibinfo{author}{V.~M. Shabaev}, and \bibinfo{author}{G.~Soff},
  \bibinfo{journal}{Phys. Rev. A} \bibinfo{volume}{\textbf{60}},
  \bibinfo{pages}{3522} (\bibinfo{date}{1999}).
\bibitem{artemyev99}
\bibinfo{author}{A.~N. Artemyev}, \bibinfo{author}{T.~Beier},
  \bibinfo{author}{G.~Plunien}, \bibinfo{author}{V.~M. Shabaev},
  \bibinfo{author}{G.~Soff}, and \bibinfo{author}{V.~A. Yerokhin},
  \bibinfo{journal}{Phys. Rev. A} \bibinfo{volume}{\textbf{60}},
  \bibinfo{pages}{45} (\bibinfo{date}{1999}).
\bibitem{gellmann51}
\bibinfo{author}{M.~Gell-Mann} and \bibinfo{author}{F.~Low},
  \bibinfo{journal}{Phys. Rev.} \bibinfo{volume}{\textbf{84}},
  \bibinfo{pages}{350} (\bibinfo{date}{1951}).
\bibitem{sucher57}
\bibinfo{author}{J.~Sucher}, \bibinfo{journal}{Phys. Rev.}
  \bibinfo{volume}{\textbf{107}}, \bibinfo{pages}{1448} (\bibinfo{date}{1957}).
\bibitem{labzowsky70}
\bibinfo{author}{{L. N. Labzowsky, Zh. Eksp. Teor. Fiz. {\bf 59}, 167 (1970)
  [Sov. Phys. JETP {\bf 32}, 94 (1970)].}}
\bibitem{braun77}
\bibinfo{author}{{M. Braun and V. Shirokov, Izv. Akad. Nauk USSR ser. fiz. {\bf
  41}, 2585 (1977) [Engl. Transl. Bull. Acad. Sci. USSR Phys. Scr. {\bf 41},
  2585 (1977)].}}
\bibitem{shabaev90}
\bibinfo{author}{{V. M. Shabaev, Teor. Mat. Fiz. {\bf 82}, 83 (1990) [Engl.
  Transl. Theor. Math. Phys. {\bf 82}, 57 (1990)].}}
\bibitem{shabaev93}
\bibinfo{author}{V.~M. Shabaev}, \bibinfo{journal}{J. Phys. B}
  \bibinfo{volume}{\textbf{26}}, \bibinfo{pages}{4703} (\bibinfo{date}{1993}).
\bibitem{shabaev02}
\bibinfo{author}{V.~M. Shabaev}, \bibinfo{journal}{Phys. Rep.}
  \bibinfo{volume}{\textbf{356}}, \bibinfo{pages}{119} (\bibinfo{date}{2002}).
\bibitem{labzowsky93karasiev}
\bibinfo{author}{L.~Labzowsky}, \bibinfo{author}{V.~Karasiev},
  \bibinfo{author}{I.~Lindgren}, \bibinfo{author}{H.~Persson}, and
  \bibinfo{author}{S.~Salomonson}, \bibinfo{journal}{Phys. Scr.}
  \bibinfo{volume}{\textbf{T46}}, \bibinfo{pages}{150} (\bibinfo{date}{1993}).
\bibitem{labzowsky02annals}
\bibinfo{author}{L.~N. Labzowsky}, \bibinfo{author}{A.~Prosorov},
  \bibinfo{author}{A.~V. Shonin}, \bibinfo{author}{I.~Bednyakov},
  \bibinfo{author}{G.~Plunien}, and \bibinfo{author}{G.~Soff},
  \bibinfo{journal}{Annals of Physics} \bibinfo{volume}{\textbf{302}},
  \bibinfo{pages}{22} (\bibinfo{date}{2002}).
\bibitem{lebigot01}
\bibinfo{author}{\mbox{\'{E}.-O}. Le~Bigot}, \bibinfo{author}{P.~Indelicato},
  and \bibinfo{author}{V.~M. Shabaev}, \bibinfo{journal}{Phys. Rev. A}
  \bibinfo{volume}{\textbf{63}}, \bibinfo{pages}{040501}
  (\bibinfo{date}{2001}).
\bibitem{lindgren01}
\bibinfo{author}{I.~Lindgren}, \bibinfo{author}{B.~\r{A}s\'en},
  \bibinfo{author}{S.~Salomonson}, and \bibinfo{author}{\mbox{A.-M}.
  {M\r{a}rtensson-Pendrill}}, \bibinfo{journal}{Phys. Rev. A}
  \bibinfo{volume}{\textbf{64}}, \bibinfo{pages}{062505}
  (\bibinfo{date}{2001}).
\bibitem{weisskopf30}
\bibinfo{author}{V.~Weisskopf} and \bibinfo{author}{E.~Wigner},
  \bibinfo{journal}{Z. Phys.} \bibinfo{volume}{\textbf{63}},
  \bibinfo{pages}{54} (\bibinfo{date}{1930}).
\bibitem{low52}
\bibinfo{author}{F.~Low}, \bibinfo{journal}{Phys. Rev.}
  \bibinfo{volume}{\textbf{88}}, \bibinfo{pages}{53} (\bibinfo{date}{1952}).
\bibitem{labzowsky83}
\bibinfo{author}{{L. N. Labzowsky, Zh. Eksp. Teor. Fiz. {\bf 85}, 869 (1983)
  [Engl. Transl. Sov. Phys. JETP {\bf 58}, 503 (1983).]}}.
\bibitem{labzowsky93only}
\bibinfo{author}{L.~N. Labzowsky}, \bibinfo{journal}{J. Phys. B}
  \bibinfo{volume}{\textbf{26}}, \bibinfo{pages}{1039} (\bibinfo{date}{1993}).
\bibitem{gorshkov89}
\bibinfo{author}{{V. G. Gorshkov, L. N. Labzowsky, and A. A. Sultanaev, Zh.
  Eksp. Teor. Fiz. {\bf 96}, 53 (1989) [Engl. Transl. Sov. Phys. JETP {\bf 69},
  28 (1989)].}}
\bibitem{karasiev92}
\bibinfo{author}{V.~V. Karasiev}, \bibinfo{author}{L.~N. Labzowsky},
  \bibinfo{author}{A.~V. Nefiodov}, \bibinfo{author}{V.~G. Gorshkov}, and
  \bibinfo{author}{A.~A. Sultanaev}, \bibinfo{journal}{Physica Scripta}
  \bibinfo{volume}{\textbf{46}}, \bibinfo{pages}{225} (\bibinfo{date}{1992}).
\bibitem{labzowsky94}
\bibinfo{author}{L.~Labzowsky}, \bibinfo{author}{V.~Karasiev}, and
  \bibinfo{author}{I.~Goidenko}, \bibinfo{journal}{J. Phys. B}
  \bibinfo{volume}{\textbf{27}}, \bibinfo{pages}{L439} (\bibinfo{date}{1994}).
\bibitem{labzowsky97goidenko}
\bibinfo{author}{L.~N. Labzowsky}, \bibinfo{author}{I.~A. Goidenko}, and
  \bibinfo{author}{D.~Liesen}, \bibinfo{journal}{Physica Scripta}
  \bibinfo{volume}{\textbf{56}}, \bibinfo{pages}{271} (\bibinfo{date}{1997}).
\bibitem{labzowsky98}
\bibinfo{author}{L.~N. Labzowsky} and \bibinfo{author}{M.~A. Tokman},
  \bibinfo{journal}{{Adv. Quant. Chem.}} \bibinfo{volume}{\textbf{30}},
  \bibinfo{pages}{393} (\bibinfo{date}{1998}).
\bibitem{labzowsky96}
\bibinfo{author}{L.~N. Labzowsky} and \bibinfo{author}{A.~O. Mitrushenkov},
  \bibinfo{journal}{Phys. Rev. A} \bibinfo{volume}{\textbf{53}},
  \bibinfo{pages}{3029} (\bibinfo{date}{1996}).
\bibitem{labzowsky96b}
\bibinfo{author}{L.~N. Labzowsky}, \bibinfo{title}{\emph{Teoriya atoma.
  Kvantovaya elektrodinamika elektronnyh obolochek i processy izlucheniya
  {\rm[}Theory of atoms. Quantum electrodynamics of the electron shells and the
  processes of radiation{\rm]} (in Russian)}} (\bibinfo{publisher}{Nauka},
  Moscow, \bibinfo{year}{1996}).
\bibitem{mohr92}
\bibinfo{author}{P.~J. Mohr}, \bibinfo{journal}{Phys. Rev. A}
  \bibinfo{volume}{\textbf{46}}, \bibinfo{pages}{4421} (\bibinfo{date}{1992}).
\bibitem{mohr93}
\bibinfo{author}{P.~J. Mohr} and \bibinfo{author}{G.~Soff},
  \bibinfo{journal}{Phys. Rev. Lett.} \bibinfo{volume}{\textbf{70}},
  \bibinfo{pages}{158} (\bibinfo{date}{1993}).
\bibitem{soff88}
\bibinfo{author}{G.~Soff} and \bibinfo{author}{P.~J. Mohr},
  \bibinfo{journal}{Phys. Rev. A} \bibinfo{volume}{\textbf{38}},
  \bibinfo{pages}{5066} (\bibinfo{date}{1988}).
\bibitem{deslattes84}
\bibinfo{author}{R.~D. Deslattes}, \bibinfo{author}{H.~F. Beyer}, and
  \bibinfo{author}{F.~Folkmann}, \bibinfo{journal}{J. Phys. B}
  \bibinfo{volume}{\textbf{17}}, \bibinfo{pages}{L689} (\bibinfo{date}{1984}).
\bibitem{briand83}
\bibinfo{author}{J.~P. Briand}, \bibinfo{author}{J.~P. Moss{\'e}},
  \bibinfo{author}{P.~Indelicato}, \bibinfo{author}{P.~Chevallier},
  \bibinfo{author}{D.~Girard-Vernhet}, \bibinfo{author}{A.~Chetioui},
  \bibinfo{author}{M.~T. Ramos}, and \bibinfo{author}{J.~P. Desclaux},
  \bibinfo{journal}{Phys. Rev. A} \bibinfo{volume}{\textbf{28}},
  \bibinfo{pages}{1413} (\bibinfo{date}{1983}).
\bibitem{briand84}
\bibinfo{author}{J.~P. Briand}, \bibinfo{author}{M.~Tavernier},
  \bibinfo{author}{R.~Marrus}, and \bibinfo{author}{J.~P. Desclaux},
  \bibinfo{journal}{Phys. Rev. A} \bibinfo{volume}{\textbf{29}},
  \bibinfo{pages}{3143} (\bibinfo{date}{1984}).
\bibitem{beiersdorfer89}
\bibinfo{author}{P.~Beiersdorfer}, \bibinfo{author}{M.~Bitter},
  \bibinfo{author}{S.~von Goeler}, and \bibinfo{author}{K.~W. Hill},
  \bibinfo{journal}{Phys. Rev. A} \bibinfo{volume}{\textbf{40}},
  \bibinfo{pages}{150} (\bibinfo{date}{1989}).
\bibitem{briand90}
\bibinfo{author}{J.~P. Briand}, \bibinfo{author}{P.~Chevallier},
  \bibinfo{author}{P.~Indelicato}, \bibinfo{author}{K.~P. Ziock}, and
  \bibinfo{author}{D.~D. Dietrich}, \bibinfo{journal}{Phys. Rev. Lett.}
  \bibinfo{volume}{\textbf{65}}, \bibinfo{pages}{2761} (\bibinfo{date}{1990}).

\end{thebibliography}

%
%
%

%
%
%
%

\newpage

%
%
%
\begin{table}
\caption{Matrix elements of the operator
$V$
for the two-electron configurations
$2 {}^1\! P_{1}$ and $2 {}^3\! P_{1}$
(eV).
The individual contributions for the Dirac-binding energies of
$2p$-electron states ($V^{(0)}$),
the one-photon exchange contribution ($V^{(1)}$)
and the two-photon contribution ($V^{(2)}$) are compiled,
respectively.
$\Eexch(2 {}^1\! P_{1})$, $\Eexch(2 {}^3\! P_{1})$
are the energies of the corresponding configurations,
where only the photon-exchange contributions are taken into account
(neglecting radiative corrections).
}
\begin{tabular}{lrrrrrrrrrr}
Contribution&$Z=\;$
10\phantom{88}&18\phantom{88}
&26\phantom{88}&30\phantom{88}&40\phantom{88}\\
\hline
\phantom{$V^{(0)}$:}~$(1s2p_{1/2}),(1s2p_{1/2})$&
$-340.7099$&$-1108.0574$&$-2325.7285$&$-3108.3193$&$-5594.0369$\\
$V^{(0)}$:~$(1s2p_{3/2}),(1s2p_{3/2})$&
$-340.2556$&$-1103.2520$&$-2304.5586$&$-3070.5057$&$-5471.5704$\\
\phantom{$V^{(0)}$:}~$(1s2p_{1/2}),(1s2p_{3/2})$&
$0$&$0$&$0$&$0$&$0$\\
\hline
\phantom{$V^{(1)}$:}~$(1s2p_{1/2}),(1s2p_{1/2})$&
$64.7130+0.0007i$&$117.2696+0.0072i$&$171.2105+0.0316i$&$198.9154+0.0560i$&$271.1021+0.1778i$\\
{$V^{(1)}$:}~$(1s2p_{3/2}),(1s2p_{3/2})$&
$67.6938-0.0007i$&$122.1610-0.0072i$&$177.1674-0.0312i$&$204.9502-0.0551i$&$275.4795-0.1724i$\\
\phantom{$V^{(1)}$:}~$(1s2p_{1/2}),(1s2p_{3/2})$&
$ 4.3418-0.0019i$&$  7.6653-0.0204i$&$ 10.7333-0.0887i$&$ 12.1369-0.1571i$&$ 15.1577-0.4954i$\\
\hline
\phantom{$V^{(2)}$:}~$(1s2p_{1/2}),(1s2p_{1/2})$&
$-2.7692$&$-2.8168$&$-2.8938$&$-2.9439$&$-3.1082$\\
{$V^{(2)}$:}~$(1s2p_{3/2}),(1s2p_{3/2})$&
$-3.5256$&$-3.5603$&$-3.6142$&$-3.6506$&$-3.7641$\\
\phantom{$V^{(2)}$:}~$(1s2p_{1/2}),(1s2p_{3/2})$&
$-1.0727$&$-1.0618$&$-1.0450$&$-1.0350$&$-1.0008$\\
\hline
\quad $\Eexch(2 {}^1\! P_{1})$&
$-273.8939$&$-981.1501$&$-2127.8323$&$-2866.5171$&$-5198.2878$\\
\quad $\Eexch(2 {}^3\! P_{1})$&
$-280.9596$&$-997.1059$&$-2160.5849$&$-2915.0368$&$-5327.6102$\\
\hline
\hline
\quad $\Delta \Eexch(2 {}^1\! P_{1})$: Appr.~1&
$ 2.1934$&$ 3.5012$&$ 3.1731$&$ 2.6890$&$ 1.5672$\\
\quad $\Delta \Eexch(2 {}^3\! P_{1})$: Appr.~1&
$-2.1934$&$-3.5012$&$-3.1731$&$-2.6890$&$-1.5672$\\
\hline
\quad $\Delta \Eexch(2 {}^1\! P_{1})$: Appr.~2&
$-0.7581$&$-0.7203$&$-0.5599$&$-0.4460$&$-0.2175$\\
\quad $\Delta \Eexch(2 {}^3\! P_{1})$: Appr.~2&
$ 0.7581$&$ 0.7203$&$ 0.5599$&$ 0.4460$&$ 0.2175$\\
\hline
\quad $\Delta \Eexch(2 {}^1\! P_{1})$: Appr.~3&
$ 0.0001$&$ 0.0019$&$ 0.0102$&$ 0.0168$&$ 0.0571$\\
\quad $\Delta \Eexch(2 {}^3\! P_{1})$: Appr.~3&
$ 0.0000$&$-0.0009$&$-0.0036$&$-0.0056$&$-0.0121$\\
\hline
\quad $\Delta \Eexch(2 {}^1\! P_{1})$: Appr.~4&
$ 0.0000$&$ 0.0002$&$ 0.0009$&$-0.0005$&$-0.0013$\\
\quad $\Delta \Eexch(2 {}^3\! P_{1})$: Appr.~4&
$ 0.0000$&$ 0.0000$&$ 0.0009$&$ 0.0018$&$ 0.0038$\\
\hline
\quad $\Delta \Eexch(2 {}^1\! P_{1})$: Appr.~5&
$ 0.0000$&$ 0.0000$&$-0.0001$&$-0.0003$&$-0.0015$\\
\quad $\Delta \Eexch(2 {}^3\! P_{1})$: Appr.~5&
$ 0.0000$&$ 0.0000$&$ 0.0001$&$ 0.0003$&$ 0.0015$\\
\hline
\hline
\quad $\Eexch(2 {}^1\! P_{1})-\Eexch(2 {}^3\! P_{1})$\\
This work&$7.0657$&$15.9557$\\
Lindgren {\it et al.} \cite{lindgren01}&$7.0657$&$15.9554$\\
\end{tabular}
\label{txquasidegx1}
\end{table}
%
%
%
%
%
%
\addtocounter{table}{-1}
\begin{table}
\caption{(\it Continued.)}
\begin{tabular}{lrrrrrrrrrr}
Contribution&$Z=\;$
50\phantom{88}&60\phantom{88}
&70\phantom{88}&80\phantom{88}&92\phantom{88}\\
\hline
\phantom{$V^{(0)}$:}~$(1s2p_{1/2}),(1s2p_{1/2})$&
$-8884.368$&$-13062.966$&$-18250.182$&$-24621.409$&$-34211.065$\\
$V^{(0)}$:~$(1s2p_{3/2}),(1s2p_{3/2})$&
$-8575.514$&$-12395.463$&$-16948.025$&$-22253.673$&$-29649.834$\\
\phantom{$V^{(0)}$:}~$(1s2p_{1/2}),(1s2p_{3/2})$&
$0$&$0$&$0$&$0$&$0$\\
\hline
\phantom{$V^{(1)}$:}~$(1s2p_{1/2}),(1s2p_{1/2})$&
$348.917+0.437i$&$434.638+0.912i$&$531.400+1.702i$&$643.793+2.930i$&$809.699+5.178i$\\
{$V^{(1)}$:}~$(1s2p_{3/2}),(1s2p_{3/2})$&
$347.960-0.415i$&$422.949-0.846i$&$501.042-1.531i$&$582.883-2.533i$&$686.996-4.210i$\\
\phantom{$V^{(1)}$:}~$(1s2p_{1/2}),(1s2p_{3/2})$&
$ 17.320-1.206i$&$ 18.430-2.491i$&$ 18.311-4.590i$&$ 16.795-7.777i$&$ 12.929-13.440i$\\
\hline
\phantom{$V^{(2)}$:}~$(1s2p_{1/2}),(1s2p_{1/2})$&
$-3.333$&$-3.635$&$-4.038$&$-4.585$&$-5.531$\\
{$V^{(2)}$:}~$(1s2p_{3/2}),(1s2p_{3/2})$&
$-3.915$&$-4.105$&$-4.339$&$-4.628$&$-5.053$\\
\phantom{$V^{(2)}$:}~$(1s2p_{1/2}),(1s2p_{3/2})$&
$-0.955$&$-0.893$&$-0.801$&$-0.771$&$-0.683$\\
\hline
\quad $\Eexch(2 {}^1\! P_{1})$&
$-8230.604$&$-11976.160$&$-16451.098$&$-21675.333$&$-28967.898$\\
\quad $\Eexch(2 {}^3\! P_{1})$&
$-8539.649$&$-12632.423$&$-17723.044$&$-23982.287$&$-33406.890$\\
\hline
\hline
\quad $\Delta \Eexch(2 {}^1\! P_{1})$: Appr.~1&
$ 0.865$&$ 0.460$&$ 0.225$&$ 0.085$&$-0.007$\\
\quad $\Delta \Eexch(2 {}^3\! P_{1})$: Appr.~1&
$-0.865$&$-0.460$&$-0.225$&$-0.085$&$ 0.007$\\
\hline
\quad $\Delta \Eexch(2 {}^1\! P_{1})$: Appr.~2&
$-0.102$&$-0.049$&$-0.023$&$-0.011$&$-0.004$\\
\quad $\Delta \Eexch(2 {}^3\! P_{1})$: Appr.~2&
$ 0.102$&$ 0.049$&$ 0.023$&$ 0.011$&$ 0.004$\\
\hline
\quad $\Delta \Eexch(2 {}^1\! P_{1})$: Appr.~3&
$ 0.153$&$ 0.358$&$ 0.759$&$ 1.461$&$ 2.931$\\
\quad $\Delta \Eexch(2 {}^3\! P_{1})$: Appr.~3&
$-0.014$&$-0.007$&$ 0.021$&$ 0.096$&$ 0.323$\\
\hline
\quad $\Delta \Eexch(2 {}^1\! P_{1})$: Appr.~4&
$-0.004$&$-0.009$&$-0.015$&$-0.035$&$-0.075$\\
\quad $\Delta \Eexch(2 {}^3\! P_{1})$: Appr.~4&
$ 0.011$&$ 0.025$&$ 0.050$&$ 0.094$&$ 0.181$\\
\hline
\quad $\Delta \Eexch(2 {}^1\! P_{1})$: Appr.~5&
$-0.004$&$-0.009$&$-0.016$&$-0.026$&$-0.041$\\
\quad $\Delta \Eexch(2 {}^3\! P_{1})$: Appr.~5&
$ 0.004$&$ 0.009$&$ 0.016$&$ 0.026$&$ 0.041$\\
\end{tabular}
\label{txquasidegx2}
\end{table}
%
%
%
%
%
%
\begin{table}
\caption{Data for the energies (in eV) of the configurations
$2 {}^1\! P_{1}$ and $2 {}^3\! P_{1}$.
Photon-exchange corrections are taken into account up to second order
in $\alpha$.
Self-energy (SE) and vacuum-polarization (VP) corrections are
taken into account only in first order.
The one-electron radiative corrections of order $\alpha^2$
,
the SE and VP screening corrections
and all the corrections of the third and higher orders are omitted.
The data are compared with the results of
Plante {\it et al.}~[2]
and
Drake~[1].
}
\begin{tabular}{lrrrrrrrrrr}
Contribution&$Z=\;$
10\phantom{88}&18\phantom{88}
&26\phantom{88}&30\phantom{88}&40\phantom{88}\\
\hline
 $E(2 {}^1\! P_{1})$, this work&
$-273.8936$&$-981.1462$&$-2127.8070$&$-2866.4666$&$-5198.0971$\\
 $E(2 {}^3\! P_{1})$, this work&
$-280.9596$&$-997.1079$&$-2160.5953$&$-2915.0545$&$-5327.6408$\\
\hline
 $E(2 {}^1\! P_{1})$, Plante {\it et al.}
&
$-273.8155$&$-981.0966$&$-2127.7712$&$-2866.4354$&$-5198.0801$\\
 $E(2 {}^3\! P_{1})$, Plante {\it et al.}
&
$-281.0132$&$-997.1451$&$-2160.6313$&$-2915.0924$&$-5327.6917$\\
\hline
 $E(2 {}^1\! P_{1})$, Drake
&
$-273.8077$&$-981.0832$&$-2127.7515$&$-2866.4129$&$-5198.0515$\\
 $E(2 {}^3\! P_{1})$, Drake
&
$-281.0054$&$-997.1303$&$-2160.6038$&$-2915.0556$&$-5327.6222$\\
\end{tabular}
\label{txquasidegx3}
\end{table}
%
%
%
\addtocounter{table}{-1}
\begin{table}
\caption{{(\it Continued.)}
}
\begin{tabular}{lrrrrrrrrrr}
Contribution&$Z=\;$
50\phantom{88}&60\phantom{88}
&70\phantom{88}&80\phantom{88}&92\phantom{88}\\
\hline
 $E(2 {}^1\! P_{1})$, this work&
$-8230.079$&$-11974.951$&$-16448.629$&$-21670.722$&$-28958.974$\\
 $E(2 {}^3\! P_{1})$, this work&
$-8539.626$&$-12632.120$&$-17721.892$&$-23979.001$&$-33397.135$\\
\hline
 $E(2 {}^1\! P_{1})$, Plante {\it et al.}
&
$-8230.078$&$-11974.972$&$-16448.682$&$-21670.812$&$-28959.135$\\
 $E(2 {}^3\! P_{1})$, Plante {\it et al.}
&
$-8539.719$&$-12632.321$&$-17722.363$&$-23980.133$&$-33400.643$\\
\hline
 $E(2 {}^1\! P_{1})$, Drake
&
$-8230.038$&$-11974.910$&$-16448.591$&$-21670.679$&$-28958.944$\\
 $E(2 {}^3\! P_{1})$, Drake
&
$-8539.592$&$-12632.088$&$-17721.944$&$-23979.373$&$-33398.993$\\
\end{tabular}
\label{txquasidegx4}
\end{table}
%
%
%
%
%
%
\begin{table}
\caption{Theoretical and experimental data for
$2 {}^3\! P_{1}-2 {}^1\! P_{1}$
transition energies (in eV).
}
\begin{tabular}{lrrrrlr}
$Z$
&This work
&Lindgren {\it et al.} \cite{lindgren01}
&Plante {\it et al.} \cite{plante94}
&Drake \cite{drake88}
&{Experiment}
&{Ref.}\\
\hline
18&$15.9617$&16.0550&$16.0485$&$16.0471$&$16.031(0.074)$
&\cite{deslattes84}\\
&&&&&$16.00(0.49)$&\cite{briand83}\\
26&$32.7883$&&$32.8601$&$32.8523$&$33.4(0.5)$&\cite{briand84}\\
&&&&&$33.23(0.45)$&\cite{beiersdorfer89,briand84}\\
92&$4438.161$&&$4441.508$&$4440.049$&$4455.(87.)$&\cite{briand90}\\
\end{tabular}
\label{txquasidegx5}
\end{table}
%
%
%

%
%
%
%
%
%

\newpage

%
%
%

%
%
%
%
%
%
\end{document}